# CORC Report TR-2006-04
# Characterizing Optimal Adword Auctions[*]

Garud Iyengar [†]    Anuj Kumar [‡]

First Version: April 2006
This version: November 14, 2006

## Abstract

We present a number of models for the adword auctions used for pricing advertising slots on search engines such as Google, Yahoo! etc. We begin with a general problem formulation which allows the privately known valuation per click to be a function of both the identity of the advertiser and the slot. We present a compact characterization of the set of all deterministic incentive compatible direct mechanisms for this model. This new characterization allows us to conclude that there are incentive compatible mechanisms for this auction with a multi-dimensional type-space that are *not* affine maximizers. Next, we discuss two interesting special cases: slot independent valuation and slot independent valuation up to a privately known slot and zero thereafter. For both of these special cases, we characterize revenue maximizing and efficiency maximizing mechanisms and show that these mechanisms can be computed with a worst case computational complexity $\mathcal{O}(n^2m^2)$ and $\mathcal{O}(n^2m^3)$ respectively, where $n$ is number of bidders and $m$ is number of slots. Next, we characterize optimal rank based allocation rules and propose a new mechanism that we call the customized rank based allocation. We report the results of a numerical study that compare the revenue and efficiency of the proposed mechanisms. The numerical results suggest that customized rank-based allocation rule is significantly superior to the rank-based allocation rules.

## 1 Introduction

Sponsored search advertising is a major source of revenue for internet search engines. Close to 98% of Google's total revenue of $6 billion for the year 2005 came from sponsored search advertisements. It is believed that more than 50% of Yahoo!'s revenue of $5.26 billion was from sponsored search advertisement. Sponsored search advertisements work as follows. A user queries a certain *adword*, i.e. a keyword relevant for advertisement, on an online search engine. The search engine returns the links to the most "relevant" webpages and, in addition, displays certain number of relevant sponsored links in certain fixed "slots" on the result page. For example, when we search for "`Delhi`" on Google, in addition to the most relevant webpages, eight sponsored links are also displayed which include links to websites of the hotels in Delhi. Every time the user clicks on any of these sponsored links, she is taken to the website of the advertiser sponsoring the link and the search engine receives certain price per click from the advertiser. It is reasonable to expect that, all things being equal, a user is more likely click on the link that is placed in a slot that is easily visible on the page. In any case, the click likelihood is a function of the slot, and therefore, advertisers have a preference over which slot carries their link and are willing to pay a higher price per click when placed on a more desirable slot. The chance that a user clicks on a sponsored link is likely to be an increasing function of the exogenous brand values of the advertisers; therefore, search engines prefer allocating more desirable slots to advertisers with higher exogenous brand value. In conclusion, the search engines need a mechanism for allocating slots to advertisers. Since auctions are very effective mechanisms for revenue generation and efficient allocation, they have become the mechanism of choice for assigning sponsored links to advertising slots.

[*]An extended abstract of this paper appeared in the *Second Workshop on Sponsored Search Auctions*, June 11, 2006 Ann Arbor, Michigan in conjunction with the *ACM Conference on Electronic Commerce (EC'06)*

[†]Industrial Engineering and Operations Research Department, Columbia University, New York, NY-10027. Email: garud@ieor.columbia.edu. Research partially supported by NSF grants CCR-00-09972, DMS-01-04282 and ONR grant N000140310514.

[‡]Industrial Engineering and Operations Research Department, Columbia University, New York, NY-10027. Email: ak2108@columbia.edu. Research partially supported by NSF grant DMS-01-04282 and ONR grant N000140310514.

Adwords auctions are dynamic in nature – the advertisers are allowed to change their bids quite frequently. In this paper, we design and analyze static models for adword auctions. We use the dominant strategy solution concept in order to ensure that the static model adequately approximates dynamic adword auctions. The main contributions of this paper are as follows.

(a) We formulate a general model for "pay-per-click" adword auctions when the privately known valuation-per-click $v_{ij}$ of advertiser $i$ is a function of the allocated slot $j$. Thus, the advertisers have a multi-dimensional type-space. We characterize the set of all dominant strategy incentive compatible, individually rational allocation rules. Using this characterization, we show that there exist incentive compatible allocation rules that are *not* affine maximizers (see Example 1). For details see §2.

(b) When private valuation-per-click $v_{ij}$ is not a function of the slot $j$, we completely solve the mechanism design problem, i.e. we characterize of the set of all dominant strategy incentive compatible, individually rational allocation rules, the unique prices that implements these rules, and the revenue maximizing mechanism and a computationally tractable implementation.

For this model, we analyze rank-based allocations rules and show how to compute an *optimal* rank based allocation rule. We show that even when the click-through-rate matrix is separable, i.e. $c_{ij}$ is of the form $c_{ij} = \phi_i \mu_j$, the efficiency maximizing rank vector and revenue maximizing rank vector are not same, moreover, the unconstrained revenue maximizing mechanism is em not rank based (see Example 2).

We also propose a new, easy to implement allocation rule that we call the customized rank based allocation rule. We show that the this new rule has significantly superior performance – both in terms of efficiency and revenue – when compared with rank based mechanisms. For details see § 3.

(c) We analyze a model in which the valuations are privately known constant up to a privately known slot and zero thereafter, henceforth called the slotted valuation model. We present two suboptimal (revenue maximizing) mechanisms for this model, which perform remarkably well on a set of synthetic data. See § 4 for details.

(d) We implement all the proposed mechanisms and test their relative performance on a set of synthetic data. See § 5.

Our models are still far from capturing all the important tradeoffs in the current business models for adword pricing. In particular, we don't model the influence of budgets, risk averseness, bidder irrationality (or bounded rationality) and diversification across adwords. There is growing literature which focuss on these aspects on adword auctions, some of which we discuss in our literature survey.

The paper is organized as follows. We discuss the relevant literature in § 1.1. In § 2 we describe the general adword auction model. In § 3 we discuss the adword auction model with slot independent private valuations. In § 4 we propose and analyze the slotted model. In § 5 we discuss the results of a preliminary numerical study and § 6 contains some concluding remarks.

## 1.1 Previous Literature

The online adword auctions have caught the attention of the academic community only recently – after all, sponsored search and the Internet itself is a recent phenomena when compared with the long tradition of research in auction theory. The papers that specifically address adword auctions from the perspective of auction theory include Aggarwal et al. (2006b), Edelman et al. (2005) and Lahaie (2006). Edelman et al. (2005) present a comprehensive introduction and history of the adword auctions. Edelman et al. (2005) study an adword auction model with slot independent valuation, separable click-though-rate and generalized second price payment rule. They observe that truth-telling is not a dominant strategy for this auction. Aggarwal et al. (2006b) also study the same model and propose a payment rule under which any *rank based* allocation rule can be made dominant strategy incentive compatible. They also show that if the click-though-rate is separable (in which case the Vickery-Clark-Groves (VCG) allocation rule is rank based) there exists an ex-post equilibrium which results in same pointwise revenue as the generalized second price auction. This equilibrium is a simple adaptation of the equilibrium in Edelman et al. (2005). We demonstrate that this revenue equivalence follows in a straightforward manner from the fact that the private information in the uniform valuation model is single dimensional. Lahaie (2006) characterizes the equilibrium bidding strategies in rank based mechanisms with first price and second price payment schemes in both complete and incomplete information setting.

Varian (2006) characterizes the Nash Equilibrium in an adword auction with first price payments. In this model, the value per click only depends on the identity of the bidder and the click through rate only depends on the slot. Varian (2006) also reports results of comparing the prices predicted by the Nash equilibrium to empirical prices.

Zhan et al. (2005), Feng (2006a) and Lim and Tang (2004) present a Bayesian Nash analysis of a related adword auction model with one dimensional private information. Feng (2006b) studies bid price cycling in online auctions. Feng et al. (2005) assume truthful bidding and study the revenue performance of alternative rank based mechanisms using simulation. Liu and Chen (2005) propose using historical bid as the prior for designing the auctions. Kitts et al. (2005) present a simplified analysis of the equilibrium behavior with very few assumptions, focussing on dynamic behavior and empirical analysis of the bid data.

Shapley and Shubik (1972) describe an efficient assignment game in which bidders are assigned to objects with each bidder receiving at most one object. See Bikhchandani and Ostroy (2006) for a recent survey of mechanisms that yield efficient equilibria for the assignment game. Leonard (1983) showed that a specific optimal dual solution of the matching linear program implements the efficiency maximizing allocation in dominant strategy.

The characterization of the incentive compatibility constraint in the slotted model is based on Iyengar and Kumar (2006) where the authors study one-sided incentives in a reverse auction model. Vohra and Malakhov (2005) also consider a similar but restrictive model. We show in § 4 that this restrictive model is not adequate in the setting considered in this paper (see Example 3). Aggarwal et al. (2006a) propose a *top-down* auction for a slotted model with bidder independent click-through-rate and show that this auction has an envy-free Nash equilibrium with the same allocation and prices as the efficiency maximizing VCG mechanism. We show in § 3 that the top-down allocation rule is a special case of the customized rank-based allocation rule.

**Notation**

We denote vectors by boldface lowercase letters, e.g. $\mathbf{v}$. A vector indexed by $-i$, (for example $\mathbf{x}_{-i}$) denotes the vector $\mathbf{x}$ with the $i$-th component excluded. We use the convention $\mathbf{x} = (x_i, \mathbf{x}_{-i})$. Scalar (resp. vector) functions are denoted by lowercase letters, e.g. $X_{ij}(\mathbf{v}_i, \mathbf{v}_{-i})$ (resp. $\mathbf{X}(\mathbf{v}_i, \mathbf{v}_{-i})$). The possible misreport of the true parameters are represented with a hat over the same variable, e.g. $\hat{\mathbf{v}}$). We will use the terms advertiser and bidder interchangeably.

## 2  Adword Auction Model

There are $n$ advertisers bidding for $m(\leq n)$ slots on a specific adword. Let $c_{ij}$ denote the click-through-rate when advertiser $i$ is assigned to slot $j$. For convenience, we will set $c_{i,m+1} = 0$ for all $i = 1, \ldots, n$.

**Assumption 1.** *The click through rates $\{c_{ij}\}$ satisfy the following conditions.*

(i) *For all bidders $i$, the rate $c_{ij}$ strictly positive and non-increasing in $j$, i.e.* all *bidders rank the slots in the same order.*

(ii) *The rates $c_{ij}$, for all $(i,j)$ pairs $i = 1, \ldots, n$, $j = 1, \ldots, m$, are known to the auctioneer.*

(iii) *Only the rates $(c_{i1}, c_{i2}, \ldots, c_{im})$ are known to bidder $i$, i.e. each bidder only knows her click-through-rates.*

The true expected per-click-value $v_{ij}$ of slot $j$ to advertiser $i$ is private information. We assume *independent private values* (IPV) setting with a commonly known prior distribution function that is continuously differentiable with density $f(\mathbf{v}_1, \ldots, \mathbf{v}_n) = \prod_{i=1}^n f_i(\mathbf{v}_i) : \mathbb{R}_+^{m \times n} \mapsto \mathbb{R}_{++}$. Note that even through we use dominant strategy as the solution concept, we still need the prior distribution in order to select the optimal mechanism. We restrict attention to direct mechanisms – the revelation principle guarantees that this does not introduce any loss of generality.

Let $\mathbf{b} \in \mathbb{R}_+^{n \times m}$ denote the bids of the $n$ bidders. An auction mechanism for this problem consists of the following two components.

1. An allocation rule $\mathbf{X} : \mathbb{R}_+^{n \times m} \mapsto \{0,1\}^{n \times m}$ that satisfies

$$\sum_{i=1}^n X_{ij}(\mathbf{b}) = 1, \quad j = 1, \ldots, m,$$
$$\sum_{j=1}^m X_{ij}(\mathbf{b}) \leq 1, \quad i = 1, \ldots, n.$$

   Thus, $\mathbf{X}(\mathbf{b})$ is a matching that matches bidders to slots as a function of the bid $\mathbf{b}$. Henceforth, we denote the set of all possible matchings of $n$ advertisers to $m$ slots by $\mathcal{M}_{nm}$.

2. A payment function $\mathbf{T} : \mathbb{R}_+^{n \times m} \mapsto \mathbb{R}^n$ that specifies what each of the $n$ bidders pay the auctioneer.

We show below that one can set the payment of the bidder who is not allocated any slot to zero without any loss of generality. Thus, we can define the per click payment $t_i$ of advertiser $i$ as

$$t_i(\mathbf{b}) = \frac{T_i(\mathbf{b})}{\sum_{j=1}^m c_{ij} X_{ij}(\mathbf{b})}.$$

For $\mathbf{v} \in \mathbb{R}_+^{n \times m}$ and $i = 1, \ldots, n$, let

$$\hat{u}_i(\mathbf{b}, \mathbf{v}; (\mathbf{X}, \mathbf{T}), \mathbf{v}_{-i}) = \sum_{j=1}^m \left( c_{ij} v_{ij} - T_i(\mathbf{b}, \mathbf{v}_{-i}) \right) X_{ij}(\mathbf{b}, \mathbf{v}_{-i}) \tag{1}$$

denote the utility the advertiser $i$ of type $\mathbf{v}_i$ who bids $\mathbf{b}$. When the mechanism $(\mathbf{X}, \mathbf{T})$ is clear by context, we will write the utility as $\hat{u}_i(\mathbf{b}, \mathbf{v}; \mathbf{v}_{-i})$.

We restrict attention to mechanisms $(\mathbf{X}, \mathbf{T})$ that satisfy the following two properties:

1. Incentive compatibility (IC): For all $\mathbf{v} \in \mathbb{R}_+^{n \times m}$ and all $i = 1, \ldots, n$,

$$\mathbf{v}_i \in \underset{\mathbf{b} \in \mathbb{R}_+^m}{\operatorname{argmax}} \{ u_i(\mathbf{b}; (\mathbf{X}, \mathbf{T}), \mathbf{v}_{-i}) \}, \tag{2}$$

   i.e. truth telling is ex-post dominant.

2. Individual rationality (IR): For all $\mathbf{v} \in \mathbb{R}_+^{n \times m}$ and all $i = 1, \ldots, n$,

$$\underset{\mathbf{b} \in \mathbb{R}_+^m}{\operatorname{argmax}} \{ u_i(\mathbf{b}; (\mathbf{X}, \mathbf{T}), \mathbf{v}_{-i}) \} \geq 0, \tag{3}$$

   i.e. we implicitly assume that the outside alternative is worth zero.

Let

$$u_i(\mathbf{v}_i; (\mathbf{X}, \mathbf{T}), \mathbf{v}_{-i}) = \max_{\mathbf{b} \in \mathbb{R}_+^m} \{ \hat{u}_i(\mathbf{b}, \mathbf{v}_i; (\mathbf{X}, \mathbf{T}), \mathbf{v}_{-i}) \} \tag{4}$$

denote the maximum attainable utility for advertiser $i$ under the mechanism $(\mathbf{X}, \mathbf{T})$. If the mechanism $(\mathbf{X}, \mathbf{T})$ is IC and IR then clearly,

$$u_i(\mathbf{v}_i; (\mathbf{X}, \mathbf{T}), \mathbf{v}_{-i}) = \hat{u}_i(\mathbf{v}_i, \mathbf{v}_i; (\mathbf{X}, \mathbf{T}), \mathbf{v}_{-i}) \tag{5}$$

Next, we develop an alternative characterization of the *IC* constraint. Fix $\mathbf{v}_{-i}$ and consider the optimization problem of $i$-th bidder,

$$u_i(\mathbf{v}_i, \mathbf{v}_{-i}) = \max_{\hat{\mathbf{v}}_i \in \mathbb{R}_+^m} \Big\{ \sum_{j=1}^m (c_{ij} v_{ij} - \mathbf{T}_i(\hat{\mathbf{v}}_i, \mathbf{v}_{-i})) X_{ij}(\hat{\mathbf{v}}_i, \mathbf{v}_{-i}) \Big\}.$$

Clearly $u_i$ is convex in $\mathbf{v}_i$ since it is a maximum of a collection of linear functions and by envelope conditions its gradient (under truth-telling) is given by

$$\nabla_{\mathbf{v}_i} u_i(\mathbf{v}_i, \mathbf{v}_{-i}) = (c_{i1} X_{i1}(\mathbf{v}_i, \mathbf{v}_{-i}), \ldots, c_{im} X_{im}(\mathbf{v}_i, \mathbf{v}_{-i}))^{\mathrm{T}} \quad a.e.$$

Thus, an incentive compatible allocation rule is always a sub-gradient of some convex function and hence integrable and monotone[1], as was first observed by Rochet (1987).

Several authors have characterized the set of incentive compatible allocation rule in quasi-linear environments (see Lavi et al. (2004); Hongwei et al. (2004); Chung and Ely (2002); Saks and Yu (2005) ) in terms of the absence of negative 2-cycles (also called weak monotonicity):

$$\sum_{j=1}^m c_{ij} \left( v_{ij} X_{ij}(\mathbf{v}) + \tilde{v}_{ij} X_{ij}(\tilde{\mathbf{v}}_i, \mathbf{v}_{-i}) \right) \geq \sum_{j=1}^m c_{ij} \left( v_{ij} X_{ij}(\tilde{\mathbf{v}}_i, \mathbf{v}_{-i}) + \tilde{v}_{ij} X_{ij}(\mathbf{v}) \right) \quad \forall \mathbf{v}_i, \tilde{\mathbf{v}}_i, \mathbf{v}_{-i},$$

i.e. the sum of utility allocated to advertiser $i$ at $\mathbf{v}_i$ and $\tilde{\mathbf{v}}_i$ under truthful bidding is greater than the sum of utility allocated to advertiser $i$ if he lies and bid $\tilde{v}_i$ at $\mathbf{v}_i$ and $v_i$ at $\tilde{\mathbf{v}}_i$. For a convex domain this condition implies that the allocation $\mathbf{X}$ is integrable and the transfer payments implementing $\mathbf{X}$ are well defined. Chung and Ely (2002) propose a new implementability condition called quasi-efficiency according to which an allocation rule $\mathbf{X}$ is *IC* if, and only if, for all $i, \theta$, there exist functions $g_i : \Theta^{n-1} \times \mathcal{A} \mapsto \mathbb{R}$ such that

$$\mathbf{X}(\theta) = \underset{a \in \mathcal{A}}{\operatorname{argmax}} \left\{ v_i(\theta_i, a) + g_i(\theta_{-i}, a) \right\},$$

where $\theta$ is the private information and $\mathcal{A}$ is the set of allocations. For efficient allocations, the functions $g_i$ is just the sum of utilities of all bidders other than $i$ at their respective type $\theta_{-i}$.

One drawback of these characterizations is that they guarantee existence of transfer payments but do not provide any way of computing them. We give a new characterization of *IC* allocation rules directly in terms of bidder dependent slot prices.

**Lemma 1.** *An allocation rule* $\mathbf{X} : \mathbb{R}_+^{n \times m} \mapsto \mathcal{M}_{nm}$ *is incentive compatible if and only if for all* $1 \leq i \leq n$ *and* $\mathbf{v}_{-i} \in \mathbb{R}_+^{(n-1) \times m}$, *there exists per-click prices* $\mathbf{p}_i \in (\mathbb{R} \cup \{\infty\})^m$ *and* $p_{i0} \in \mathbb{R} \cup \{\infty\}$ *such that,*

$$X_{ij}(\mathbf{v}) = 1 \Rightarrow c_{ij}(v_{ij} - p_{ij}) \geq \max \left\{ \{c_{ik}(v_{ik} - p_{ik})\}_{k=1}^m, -p_{i0} \right\}.$$

**Proof:** Fix $i$, $\mathbf{v}_{-i}$ and suppress the dependence on $\mathbf{v}_{-i}$. $\mathbf{X}$ is IC iff there exists convex functions (i.e. the indirect utilities) $u_i : \mathbb{R}_+^m \mapsto \mathbb{R}$ such that

$$\nabla u_i(\mathbf{v}_i) = c_{ij} \mathbf{e}_j \text{ for some } j \in \{0, 1, \ldots, m\} \quad a.e.$$

---
[1] The function $\mathbf{X} : \mathbb{R}_+^m \mapsto \mathbb{R}^m$ is monotone if for every $\mathbf{y}, \mathbf{z} \in \mathbb{R}_+^m$, $(\mathbf{y} - \mathbf{z})^{\mathrm{T}}(\mathbf{X}(\mathbf{y}) - \mathbf{X}(\mathbf{z})) \geq 0$

where $\mathbf{e}_j$ is the $j$-th unit vector, $\mathbf{e}_0 = \mathbf{0}$ and $c_{i0} = 0$. Since a convex function is absolutely continuous, it follows that $u_i$ is piecewise linear. Furthermore, a piecewise linear function is convex if, and only if, it is the pointwise maximum of each of its pieces; thus,

$$u_i(\mathbf{v}_i) = \max_{0 \leq j \leq m} \left\{ c_{ij} \mathbf{e}_j^\mathrm{T} \mathbf{v}_i - c_{ij} p_{ij} \right\} \quad \forall \mathbf{v}_i. \tag{6}$$

Define

$$\mathcal{S}_j = \{\mathbf{v} \in \mathbb{R}_+^m | X_{ij}(\mathbf{v}) = 1\}, \ j = 1, \ldots, m, \quad \text{and} \quad \mathcal{S}_0 = \{\mathbf{v} \in \mathbb{R}_+^m | X_{ij}(\mathbf{v}) = 0 \ \forall 1 \leq j \leq m\}.$$

Recall that $u_i(v_i) = \sum_{j=1}^m (c_{ij} v_{ij} - T_i(v_i)) X_{ij}(\mathbf{v}_i)$. We claim that $T_i(v_i) = T_{ij}$ for all $v_i \in \mathcal{S}_j$, i.e. the payment for bidder $i$ does not change with her bid as long as she gets the same slot[2]. Suppose this is not the case and there exists $\mathbf{v}_i^1 \neq \mathbf{v}_i^2 \in \mathcal{S}_j$ such that $T_i(\mathbf{v}_i^1) < T_i(\mathbf{v}_i^2)$. Then the bidder with valuations $\mathbf{v}_i^2$ would lie and bid $\mathbf{v}_i^1$. Thus,

$$u_i(v_i) = \sum_{j=1}^m (c_{ij} v_{ij} - T_{ij}) X_{ij}(\mathbf{v}_i) \tag{7}$$

Comparing (6) and (7), and noting that $X_{ij}(\mathbf{v}) = 1$ iff $\nabla u_i = c_{ij} e_j$, we get $T_{ij} = x_{ij} p_{ij}$ and

$$\mathbf{X}_i(\mathbf{v}_i) \in \operatorname*{argmax}_{0 \leq j \leq m} \{c_{ij}(v_{ij} - p_{ij})\}$$

where we set $v_{i0} = 0$ for notational ease. This establishes the result. ∎

Since $-p_{i0}$ is the surplus of bidder $i$ when she is not assigned any slot, $IR$ implies that $p_{i0} \leq 0$. For any given $IC$ and $IR$ mechanism and a fixed $\mathbf{v}_{-i}$, let $\underline{p} = \min_{0 \leq j \leq m}(c_{ij} p_{ij}) < 0$ then $\tilde{p}_{ij} = p_{ij} - \underline{p}/c_{ij}$ also satisfies $IC$. To see that $\tilde{\mathbf{p}}$ satisfies $IR$, let $j^* \in \operatorname{argmax}_{0 \leq j \leq m}\{c_{ij}(v_{ij} - p_{ij})\}$. Then

$$\begin{aligned} c_{ij^*}(v_{ij^*} - p_{ij^*}) &\geq c_{ik}(v_{ik} - p_{ik}), &\forall k \neq j^* \\ &\geq -c_{ik} p_{ik}, &\forall k \neq j^* \\ &\geq -\underline{p}; \end{aligned}$$

implying $c_{ij^*}(v_{ij^*} - \tilde{p}_{ij^*}) \geq 0$. Since the auctioneer would always like to minimize the bidder surplus, in the remainder of the paper, we will assume that all the prices are restricted to be positive and $p_{i0}$ is set to 0.

The above lemma can be interpreted as follows. An allocation rule, $\mathbf{X}$ is $IC$ if, and only if, there exist bidder-dependent slot prices such that bidders self-select the slot allocated to them by $\mathbf{X}$. Thus, any deterministic incentively compatible mechanism is uniquely identified by the pricing rules $p_i : \mathbb{R}_+^{(n-1) \times m} \mapsto (\mathbb{R} \cup \{\infty\})^m$.

To further understand the relationship between the characterization in Lemma 1 and integrability and monotonicity of the $IC$ allocation rule. Fix $\mathbf{v}_{-i}$. Lemma 1 implies an allocation rule is $IC$ if and only if the following two conditions are satisfied.

(1) If we increase $v_{ij}$, keeping the rest of the component of $\mathbf{v}_i$ constant, there exist a threshold value $p_{ij}$ such that for all $v_{ij} \leq p_{ij}$, advertiser $i$ is not allocated a slot $j$ and for all $v_{ij} > p_{ij}$ advertiser $i$ is allocated slot $j$.

---
[2]This claim has been observed in several previous works in more general settings.

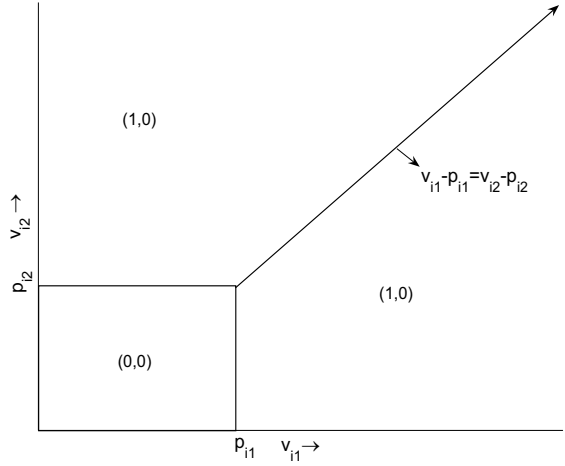

Figure 1: *IC* Allocation Rules

(2) The normal to the hyperplane separating the region in which advertiser $i$ is allocated slot $j$ (i.e. $S_j$) and the region in which advertiser $i$ allocated slot $k$ (i.e. $S_k$) is parallel to $(0, \ldots, c_{ij}, \ldots, -c_{ik}, \ldots, 0)$.

Condition (1) implies monotonicity of $\mathbf{X}$, i.e. the convexity of the surplus function, and condition (2) implies integrability of $\mathbf{X}$. Figure 1 illustrates the above conditions when there are only two advertising slots. For fixed $\mathbf{v}_{-i}$, there exist $p_{i1} \geq 0$ and $p_{i2} \geq 0$ such that

$$\mathbf{X}_i = \begin{cases} (0,0) & v_{i1} \leq p_{i1}, v_{i2} \leq p_{i2} \\ (1,0) & v_{i1} \geq p_1, c_{i2}(v_{i2} - p_{i2}) \leq c_{i1}(v_{i1} - p_{i1}) \\ (0,1) & \text{otherwise} \end{cases}$$

## 2.1 Social surplus maximization

It is well known that truth telling can be implemented in dominant strategies by the Vickery-Clark-Groves (VCG) mechanism (Groves (1979)) using the allocation rule $\mathbf{X}^e$ that maximizes the social surplus

$$\Phi(\mathbf{v}, n) = \max_{\mathbf{X} \in \mathcal{M}_{nm}} \sum_{i=1}^{n} \sum_{j=1}^{m} c_{ij} v_{ij} x_{ij} \tag{8}$$

It is also well known that the linear programming (LP) relaxation of (8) obtained by relaxing the constraint $X_{ij} \in \{0,1\}$ to $X_{ij} \in [0,1]$ is equivalent to a maximum weight flow problem on bipartite graph with unit capacities. For any such network flow problem, there exists an optimal flow that takes values in the set $\{0,1\}$, i.e. the flow is optimal for (8), and can be computed $\mathcal{O}(nm^2)$ time (see, e.g. Ahuja et al. (1993)).

Let

$$p_i^e(\mathbf{v}) = \sum_{j=1}^{m} \frac{1}{c_{ij}} \Big\{ v_{ij} - (\Phi(\mathbf{v}, n) - \Phi(\mathbf{v}_{-i}, n-1)) \Big\} X_{ij}(\mathbf{v}) \tag{9}$$

denote the per-click VCG prices. For $j = 1, \ldots, m$, let $i_j^*$ denote the index of the bidder assigned to slot $j$, i.e. $X_{i_j^*, j} = 1$. Leonard (1983) established that the vector of slot prices $\nu_j^e = \sum_{i=1}^n p_i^e(\mathbf{v}) X_{ij}(\mathbf{v})$, $j = 1, \ldots, m$, are the unique optimal solution of the LP

$$\begin{aligned} \min \quad & \sum_{j=1}^m \nu_j \\ \text{s.t.} \quad & \rho_i + \nu_j \geq c_{ij} v_{ij}, && i = 1, \ldots, n, j = 1, \ldots, m, \\ & \rho_{i_j^*} + \nu_j = c_{i_j^* j} v_{i_j^* j}, && j = 1, \ldots, m, \\ & \boldsymbol{\rho}, \boldsymbol{\nu} \geq \mathbf{0}. \end{aligned}$$

Since $\mathbf{X}^e(\mathbf{v})$ is a piecewise constant function of $\mathbf{v}$ and given $\mathbf{X}^e$, $\boldsymbol{\nu}^e(\mathbf{v})$ is a piecewise linear function of $\mathbf{v}$, it follows that the per-click VCG price $p_i^e(\mathbf{v})$ is a piecewise linear function of $\mathbf{v}$.

Let

$$\tilde{\Phi}_j(\mathbf{v}_{-i}) = \max \Big\{ \sum_{k=1, k\neq i}^n \sum_{l=1, l\neq j}^m c_{kl} v_{kl} x_{kl} : x_{ij} = 1, \mathbf{X} \in \mathcal{M}_{nm} \Big\},$$

i.e. $\tilde{\Phi}_j(\mathbf{v}_{-i})$ denotes the maximum achievable social surplus excluding the bidder $i$ and the slot $j$. Then the bidder dependent slot prices defined in Lemma 1 are given by

$$p_{ij}(\mathbf{v}_{-i}) = \frac{1}{c_{ij}} \Big( \Phi(\mathbf{v}_{-i}, n-1) - \tilde{\Phi}_j(\mathbf{v}_{-i}) \Big). \tag{10}$$

It is easy to check that if bidder $i$ is assigned slot $j$ in $\mathbf{X}^e$, the price $p_{ij}(\mathbf{v}_{-i}) = p_i^e(\mathbf{v})$. From (10) it is clear that the prices $\mathbf{p}_i(\mathbf{v}_{-i})$ are piece-wise linear functions of $\mathbf{v}_{-i}$. These prices can be efficiently computed by solving the two optimal weighted matching problems.

## 2.2 Revenue maximization

We first consider the case $n = 1$. This corresponds to monopoly pricing of stationary advertisement, e.g. lease pricing of slots on the public webpages. It follows from Lemma 1 that a revenue maximizing mechanism for a risk-neutral auctioneer is to allow the bidders to self-select slots based on the posted slot prices

$$\mathbf{p}^* \in \underset{\mathbf{p} \in \mathbb{R}_+^m}{\operatorname{argmax}} \sum_{j=1}^m p_j c_j \mathbb{E}\Big[ \mathbf{1}\Big( j \in \underset{1 \leq k \leq m}{\operatorname{argmax}}\{c_k(v_k - p_k)\}, v_j \geq p_j \Big) \Big], \tag{11}$$

where $\mathbb{E}$ denotes the expectation with respect to the prior distribution $f(\mathbf{v})$ of the valuation vector $\mathbf{v} = (v_1, \ldots, v_m)$.

For $n > 1$, the optimal revenue optimizing mechanism is the solution of the stochastic optimization problem

$$\begin{aligned} \max \quad & \mathbb{E}\Big[ \sum_{i=1}^n \sum_{j=1}^m c_{ij} p_{ij}(\mathbf{v}_{-i}) X_{ij}(\mathbf{v}) \Big] \\ \text{s.t.} \quad & \mathbf{X}(\mathbf{v}) \in \mathcal{M}_{nm} \quad \text{a.s.} \\ & v_{ij} - p_{ij}(\mathbf{v}) = \max_k \{v_{ik} - p_{ik}(\mathbf{v}_{-i})\}, \quad \forall \mathbf{v}, i, j \text{ s.t. } x_{ij}(\mathbf{v}) = 1. \end{aligned} \tag{12}$$

The stochastic optimization problem (12) is likely to be computationally hard and very sensitive to the prior distribution. See Rochet and Chone (1998) for a general treatment of the mechanism design problem with multidimensional type.

## 2.3 Affine maximizers vs general pricing rules

In this section we relate the bidder-dependent per-click slot prices to prices implied by allocation rules that maximize an affine function of the bidder surplus.

In our setting, a mechanism $(\mathbf{X}(\mathbf{v}), \mathbf{T}(\mathbf{v}))$ is called an affine maximizer auction if there exists constants $\{w_{ij}\}$ and $\{r_{ij}\}$ such that $\mathbf{X}(\mathbf{v})$ maximizes

$$\Phi^{w,r}(\mathbf{v}) = \max_{\mathbf{X} \in \mathcal{M}_{nm}} \sum_{i=1}^{n} \sum_{j=1}^{m} [w_i c_{ij} v_{ij} + r_{ij}] X_{ij},$$

and the payment $\mathbf{T}_i^{w,r}(\mathbf{v})$ when bidder $i$ is assigned to slot $j$ is given by

$$\mathbf{T}_i^{w,r}(\mathbf{v}) = \frac{1}{w_i} \Big[ \Phi^{w,r}(\mathbf{v}_{-i}, n-1) - \Phi^{w,r}(\mathbf{v}) + (w_i c_{ij} v_{ij} + r_{ij}) \Big]$$

The constants $\{r_{ij}\}$ can be interpreted as bidder-dependent slot reservation prices. The affine maximizer allocation rule $\mathbf{X}(\mathbf{v})$ also corresponds to an optimal flow in an appropriately defined network flow problem and the payments $\mathbf{T}(\mathbf{v})$ correspond to an appropriately defined minimal dual optimal vector.

The bidder-dependent slot prices $p_{ij}(\mathbf{v}_{-i})$ that implement the affine-maximizer allocation rule are given by

$$p_{ij}(\mathbf{v}_{-i}) = \frac{1}{w_i} \Big[ \Phi^{w,r}(\mathbf{v}_{-i}, n-1) - \tilde{\Phi}_j^{w,r}(\mathbf{v}) \Big],$$

where

$$\tilde{\Phi}_j(\mathbf{v}_{-i}) = \max \Big\{ \sum_{i=1}^{n} \sum_{k=1}^{m} (w_i c_{ik} v_{ik} + r_{ik}) x_{ik} : x_{ij} = 1, \mathbf{X} \in \mathcal{M}_{nm} \Big\}.$$

Since $\Phi^{w,r}(\mathbf{v}_{-i}, n-1)$ and $\tilde{\Phi}_j^{w,r}(\mathbf{v}_{-i})$ are piece-wise linear, the prices $p_{ij}(\mathbf{v}_{-i})$ are piecewise linear functions of $\mathbf{v}$.

Roberts (1979) showed that in a quasi-linear preference domain for a large enough type space (in particular, when the type space is $\mathbb{R}^{|\mathcal{A}|}$ where $\mathcal{A}$ is the allocation space) affine maximizers are the *only* dominant strategy implementable allocation rules. Given this result, a natural question that arises is whether there exist IC allocation rules in a quasi-linear environment with a given type space that are *not* affine-maximizers. Lavi et al. (2004) raise this question for a matching problem which does not satisfy the *conflicting preferences* constraint (see Open Problem 2, page 36). Since Lemma 1 does not restrict the bidder-dependent prices $\mathbf{p}_i(\mathbf{v}_{-i})$ to be of a particular form, whereas the bidder-dependent prices corresponding to affine-maximizers are piecewise linear function of $\mathbf{v}$, there is a possibility that affine-maximizers are a strict subset of IC allocation rules.

**Example 1.** Consider an adword auction with two slots and two bidders. Let $\mathbf{X}(\mathbf{v})$ denote the allocation rule that allocates slot 1 to bidder 1 if, and only if,

$$v_{11} - v_{12} \geq \text{sign}(v_{21} - v_{22}) \cdot (\|v_{21} - v_{22}\|)^{1+\gamma}$$

for some $0 < \gamma < 1$ and to bidder 2 otherwise, and assigns slot 2 to the unassigned bidder.

It is easy to check that this allocation rule is $IC$ with prices

$$\begin{aligned} p_{11}(\mathbf{v}_2) = -p_{12}(\mathbf{v}_2) &= \tfrac{1}{2} \cdot \text{sign}(v_{21} - v_{22}) \cdot (\|v_{21} - v_{22}\|)^{1+\gamma}, \\ p_{21}(\mathbf{v}_1) = -p_{22}(\mathbf{v}_1) &= \tfrac{1}{2} \cdot \text{sign}(v_{11} - v_{12}) \cdot (\|v_{11} - v_{12}\|)^{\frac{1}{1+\gamma}}, \end{aligned}$$

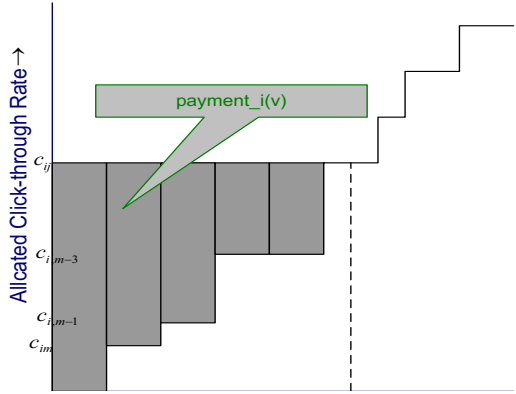

Figure 2: Incentive Compatibility: Allocated click-though-rate as a function of valuation

where the last expression follows from the fact that slot 1 is assigned to bidder 2 whenever $v_{21} - v_{22} > \text{sign}(v_{11} - v_{12}) \cdot (\|v_{11} - v_{12}\|)^{\frac{1}{1+\gamma}}$.

The prices $\{p_{ij}(\mathbf{v}_{-i})\}$ are strictly non-linear functions of $\mathbf{v}$. Consequently, $\mathbf{X}$ is an IC allocation rule that is *not* an affine maximizer. ∎

We conclude this section with the following theorem.

**Theorem 1.** *There exist individually rational and incentive compatible deterministic direct mechanism which are not affine maximizers.*

## 3 Slot independent valuations

In this section, we consider the special case where the true per-click valuations of all the bidders are slot independent, i.e. $v_{ij} = v_i$ for all $i = 1, \ldots, n$, $j = 1, \ldots, m$. Thus, the type-space of the bidders is single-dimensional.

### 3.1 Incentive compatible mechanisms

**Lemma 2.** *The following are equivalent characterizations of IC allocation rules.*

(a) *The click-through-rate $\sum_{j=1}^{m} c_{ij} X_{ij}(v_i, \mathbf{v}_{-i})$ is non-decreasing in $v_i$ for all fixed $\mathbf{v}_{-i}$.*

(b) *For all $i$ and $\mathbf{v}_{-i}$ there exist thresholds $0 \leq a_{im}(\mathbf{v}_{-i}) \leq a_{i,m-1}(\mathbf{v}_{-i}) \leq \cdots \leq a_{i,1}(\mathbf{v}_{-i}) \leq \infty$ such that bidder $i$ is assigned slot $j$ iff $v_i \in (a_{ij}(\mathbf{v}_{-i}), a_{i,j+1}(\mathbf{v}_{-i})]$.*

(c) *For all $i$ and $\mathbf{v}_{-i}$, there exist slot prices $p_{ij}(\mathbf{v}_{-i})$ of the form*

$$p_{ij}(\mathbf{v}_{-i}) = \frac{1}{c_{ij}} \sum_{k=j}^{m} \left(a_{ik}(\mathbf{v}_{-i}) - a_{i,k+1}(\mathbf{v}_{-i})\right)(c_{ij} - c_{i,k+1}), \tag{13}$$

where $0 \leq a_{im}(\mathbf{v}_{-i}) \leq a_{i,m-1}(\mathbf{v}_{-i}) \leq \cdots \leq a_{i,1}(\mathbf{v}_{-i}) \leq \infty$ such that bidders self select the slot assigned to them.

**Proof:** Part (a) follows immediately from Holmstrom's Lemma (see, p.70 in Milgrom (2004)).

Since $X_{ij} \in \{0,1\}$ and $c_{ij} \geq c_{i,j+1}$ by Assumption 1(i) the total click-through-rate $\sum_{j=1}^{m} c_{ij} X_{ij}(v_i, \mathbf{v}_{-i})$ is non-decreasing with $v_i$ if, and only if, there exist thresholds $0 \leq a_{im} \leq a_{i,m-1} \leq \cdots \leq a_{i,1} \leq \infty$ such that the allocation rule $\mathbf{X}(\mathbf{v})$ allocates slot $j$ to advertiser $i$ iff $v_i \in (a_{ij}, a_{i,j+1}]$. This is illustrated in Figure 2.

It is easy to check that prices of the form (13) results in bidder $i$ self-selecting slot $j$ if the valuation $v_i \in (a_{ij}(\mathbf{v}_{-i}), a_{i,j+1}(\mathbf{v}_{-i})]$. Thus, part (b) implies that the prices result in an IC allocation rule.

To prove the converse, observe that the payment $T_i(v_i; \mathbf{v}_{-i})$ under any IC allocation rule $\mathbf{X}$ must be of the form

$$T_i(v_i; \mathbf{v}_{-i}) = \sum_{j=1}^{m} c_{ij} v_i X_{ij}(v_i, \mathbf{v}_{-i}) - \int_0^{v_i} \Big( \sum_{j=1}^{m} c_{ij} X_{ij}(u, \mathbf{v}_{-i}) \Big) du - u_i(0, \mathbf{v}_{-i}).$$

It is easy to check that $T_i(v_i; \mathbf{v}_{-i}) + u_i(0, \mathbf{v}_{-i})$ is equal to the area of the shaded region in Figure 2. Thus per-click price for slot $j = 1, \ldots, m$,

$$\begin{aligned}
p_{ij}(\mathbf{v}_{-i}) &= \frac{T_i(v_i; \mathbf{v}_{-i}) + u_i(0; \mathbf{v}_{-i})}{c_{ij}}, \\
&= \frac{1}{c_{ij}} \Big\{ (a_{im} - 0)(c_{ij} - 0) + (a_{i,m-1} - a_{im})(c_{ij} - c_{i,m-1}) + \cdots + (a_{ij} - a_{i,j+1})(c_{ij} - c_{i,j+1}) \Big\} \\
&= \frac{1}{c_{ij}} \sum_{k=j}^{m} (a_{ik} - a_{i,k+1})(c_{ij} - c_{i,k+1}),
\end{aligned}$$

where we have set $c_{i0} = 0$ for all $i = 1, \ldots, n$. ∎

Note that by interchanging the order of integration, i.e. by computing the area of "horizontal" rectangles in Figure 2, the prices $p_{ij}(\mathbf{v}_{-i})$ can be alternatively written as

$$p_{ij}(\mathbf{v}_{-i}) = \frac{1}{c_{ij}} \sum_{k=j}^{m} (c_{ik} - c_{i,k+1}) a_{ij}. \tag{14}$$

### 3.2 Revenue maximization

From Myerson (1981), it follows that expected revenue of the auctioneer under any dominant strategy incentive compatible allocation rule $\mathbf{X}$ is given by

$$\mathbb{E}\left[ \sum_{i=1}^{n} \sum_{j=1}^{m} c_{ij} \left( v_i - \frac{1 - F_i(v_i)}{f_i(v_i)} \right) X_{ij}(\mathbf{v}) + \sum_{i=1}^{n} u_i(0, \mathbf{v}_{-i}) \right]$$

where $f_i : \mathbb{R}_+ \mapsto \mathbb{R}_{++}$ is the prior density of $v_i$, $i = 1, \ldots, n$. Thus, any two mechanisms (direct or indirect) which at a dominant strategy equilibrium agree on the point wise assignment $\mathbf{X}(\mathbf{v})$ for every $\mathbf{v}$ and on the equilibrium utilities $u_i(0, \mathbf{v}_{-i})$ for all $i, \mathbf{v}_{-i}$ result in identical expected revenues. Aggarwal et al. (2006b); Edelman et al. (2005) explicitly construct one equilibrium for the *generalized second price* auction which results in same *pointwise* revenue as the truth-telling equilibrium in the direct mechanism.

Let
$$\mathbf{X}^*(\mathbf{v}) \in \underset{\mathbf{X} \in \mathcal{M}_{nm}}{\operatorname{argmax}} \left\{ \sum_{i=1}^{n} \sum_{j=1}^{m} c_{ij} \left( v_i - \frac{1 - F_i(v_i)}{f_i(v_i)} \right) X_{ij} \right\}.$$

Suppose the virtual valuations per click $\nu_i(v_i) = v_i - \frac{1-F_i(v_i)}{f_i(v_i)}$ are non-decreasing. Then the allocation rule $\mathbf{X}^*$ results in a non-decreasing total-click-through for each of the bidders. Hence, part (a) in Lemma 2 implies that $\mathbf{X}^*$ is IC. Since the pointwise maximum is an upper bound on any expected revenue maximizing allocation, $(\mathbf{X}^*, \mathbf{T}^*)$ is expected revenue maximizing, dominant strategy IC, IR rational allocation rule with per-click prices given by (13). When the virtual valuations $\nu(v_i)$ are non-monotonic, the revenue maximizing mechanism can be constructed by first *ironing* (see Myerson (1981)) the virtual valuation to obtain a non-decreasing virtual valuations $\tilde{\nu}_i(v_i)$ and then using the construction above.

In the rest of this section, we show how to efficiently compute the payments, or equivalently, thresholds corresponding to the rule $\mathbf{X}^*$. Consider the allocation rules of the form,
$$\mathbf{X}^{\psi}(\mathbf{v}) \in \underset{\mathbf{X} \in \mathcal{M}_{nm}}{\operatorname{argmax}} \left\{ \sum_{i=1}^{n} \sum_{j=1}^{m} c_{ij} \psi_i(v_i) X_{ij} \right\}$$

for any set of $\psi_i : \mathbb{R}_+ \mapsto \mathbb{R}_+$, $\psi_i \in \mathcal{C}[\mathbb{R}]$ and $\psi_i$ non-decreasing. We call $\mathbf{X}^{\psi}$ *monotone maximizer* with respect to the set of monotone transformations $\psi_i, i = 1, \ldots, n$. Given $\mathbf{v}$, $\mathbf{X}^{\psi}(\mathbf{v})$ can computed efficiently in $\mathcal{O}(m^2 n)$ time as a solution to optimal weighted matching problem. It is straight forward to observe that $\mathbf{X}^{\psi}$ is incentive compatible and hence has the form presented in Figure 2.

As a first step towards computing the slot prices that implement the allocation rule $\mathbf{X}^{\psi}$ we solve the parametric assignment problem

$$\max_{\mathbf{X} \in \mathcal{M}_{nm}} \left\{ \lambda \sum_{j=1}^{m} c_{i_0 j} X_{i_0 j} + \sum_{k=1, k \neq i}^{n} \sum_{j=1}^{m} v_k c_{kj} X_{kj} \right\}, \tag{15}$$

where $\lambda$ is the parameter. It is clear that (15) is equivalent to the parametric minimum cost network flow problem

$$\begin{aligned}
\text{minimize} \quad & -\lambda \sum_{j=1}^{m} c_{i_0 j} X_{i_0 j} - \sum_{k=1, k \neq i}^{n} \sum_{j=1}^{m} v_k c_{kj} X_{kj} \\
\text{subject to} \quad & \sum_{j=1}^{m} X_{ij} - X_{si} = 0 \quad \forall i, \\
& \sum_{i=1}^{n} X_{ij} = 1 \quad \forall j, \\
& \sum_{i=1}^{n} X_{si} = m
\end{aligned} \tag{16}$$

on a graph $\mathcal{G}$ defined as follows.

(i) $\mathcal{G}$ has one node for every bidder and slot, and one additional node $s$.

(ii) $\mathcal{G}$ has unit capacity directed arcs from each bidder $k \neq i$ to each slot $l$ with cost $-v_k c_{kl}$ and from $s$ to each bidder with cost zero.

**Algorithm 1** OPTMATCH($i_0, \mathbf{v}_{-i_0}, c$)

1: $\mathbf{a} \leftarrow \infty, \overline{\lambda} = 0$.
2: Solver optimal matching problem at $\lambda = 0$.
3: $(\mathbf{T}, \mathbf{L}, \mathbf{U}) \leftarrow$ optimal spanning tree structure for $\lambda = 0$.
4: **while** $\overline{\lambda} < \infty$ **do**
5:    $\pi^1 \leftarrow$ optimal node potentials at tree $\mathbf{T}$ for $\lambda = 1$ and arc costs equal to zero for all arcs $(k, l)$ with $k \neq i_0$ and $\pi^1(s) \leftarrow 0$.
6:    $\pi^0 \leftarrow$ optimal node potentials at tree $\mathbf{T}$ for $\lambda = 0$ and $\pi^0(s) \leftarrow 0$
7:    $\overline{\lambda} \leftarrow \min \left\{ \frac{\pi_j^0 - \pi_{i_0}^0}{c_{i_0 j} - \pi_j^1 + \pi_{i_0}^1} : (i_0, j) \notin \mathbf{T}, c_{i_0 j} - \pi_j^1 + \pi_{i_0}^1 \geq 0 \right\}$
8:    $(i_0, j^*) \leftarrow$ the edge achieving the minimum in the previous step.
9:    $a_{i_0, j^*} \leftarrow \overline{\lambda}$
10:   Perform a network-simplex pivot with $(i_0, j^*)$ as the entering arc.
11:   $(\mathbf{T}, \mathbf{L}, \mathbf{U}) \leftarrow$ optimal spanning tree structure for $\lambda = \overline{\lambda}$.
12: **end while**

(iii) $\mathcal{G}$ has unit capacity directed arcs from bidder $i$ to each slot $j$ with cost $-\lambda c_{ij}$.

(iv) Each of the slots has unit demand and the node $s$ has a supply of $m$ units.

**Lemma 3.** OPTMATCH *correctly computes the thresholds $a_{i_0 j}$ that which bidder $i_0$ is assigned to slot $j = 1, \ldots, m$, and the worst case running time of the algorithm is $\mathcal{O}(m^2 n)$.*

**Proof:** Given a spanning tree structure $(\mathbf{T}, \mathbf{L}, \mathbf{U})$, the node potential $\pi^\lambda$ corresponding to the parameter value $\lambda$ is given by $\pi^\lambda = \pi^0 + \lambda \pi^1$ with $\pi^0$ and $\pi^1$ as computed in step 5 and 6. At each optimal spanning tree structure, the non-negativity of reduced costs for each parametric edge $(i_0, j)$ gives a bound on $\lambda$ as follows:

$$\begin{aligned}
&-\lambda c_{i_0, j} - (\pi_{i_0}^0 + \lambda \pi_{i_0}^1) + (\pi_j^0 + \lambda \pi_j^1) \geq 0 \\
&\iff \lambda(c_{i_0, j} - \pi_j^1 + \pi_{i_0}^1) \leq (\pi_j^0 - \pi_{i_0}^0) \\
&\iff \begin{cases} \lambda \leq \frac{(\pi_j^0 - \pi_{i_0}^0)}{(c_{i_0, j} - \pi_j^1 + \pi_{i_0}^1)}, & c_{i_0, j} - \pi_j^1 + \pi_{i_0}^1 \geq 0, \\ \lambda \geq \frac{(\pi_j^0 - \pi_{i_0}^0)}{(c_{i_0, j} - \pi_j^1 + \pi_{i_0}^1)}, & c_{i_0, j} - \pi_j^1 + \pi_{i_0}^1 \leq 0. \end{cases}
\end{aligned} \quad (17)$$

Since $\pi_j^1 \geq \pi_k^1$ for each non-parametric edge $(k, j)$, the reduced cost of non-parametric edges, $-v_k c_{kj} - \pi_i^0 + \pi_j^0 + \lambda(\pi_j^1 - \pi_k^1)$ is positive for all values of $\lambda$ greater than the current $\lambda$. Thus, the threshold on $\overline{\lambda}$ up to which the current spanning tree structure remains optimal is equal to the minimum of the upper bounds in (17).

After the initial solution, the algorithm performs at most $m$ pivots each costing $\mathcal{O}(mn)$, each returning a threshold at which advertiser $i$ can be assigned slot $j$. Thus, the overall complexity of OPTMATCH is $\mathcal{O}(m^2 n)$. ∎

This proof is adapted from the solution of exercise 11.48 in Ahuja et al. (1993).

Next we use OPTMATCH to compute slot prices implementing the monotone maximizer allocation $\mathbf{X}^\psi$.

**Algorithm 2** COMPUTEPRICES

1: $\mathbf{z} \leftarrow (\psi_1(v_1), \ldots, \psi_n(v_n))$, $c_{i,m+1} \leftarrow 0, a_{i,m+1} \leftarrow 0 \quad \forall i$.
2: **for** i=1 to n **do**
3: $\quad \tilde{\mathbf{a}} \leftarrow \text{OPTMATCH}(i, \mathbf{z}_{-i}, \mathbf{c})$.
4: $\quad$ **for** $j = 1$ to $m$ **do**
5: $\quad\quad a_{ij} \leftarrow \psi_i^{-1}(\tilde{a}_{ij})$
6: $\quad\quad$ **if** $a_{ij} = \infty$ **then**
7: $\quad\quad\quad a_{ij} \leftarrow \min_{j<k\leq m} a_{ik}$
8: $\quad\quad$ **end if**
9: $\quad$ **end for**
10: $\quad$ **for** j=1 to m **do**
11: $\quad\quad p_{ij} \leftarrow \frac{1}{c_{ij}} \sum_{k=j}^{m}(a_{ik} - a_{i,k+1})(c_{ij} - c_{i,k+1})$
12: $\quad$ **end for**
13: **end for**

**Lemma 4.** *Algorithm* COMPUTEPRICES *correctly computes the prices $p_{ij}$ implementing $\mathbf{X}^\psi$ in $\mathcal{O}(n^2 m^2)$ time.*

**Proof:** Without loss of generality, consider the computations for bidder 1. The call to the OPT-MATCH returns the thresholds at which bidder 1 get slot $j$ in the virtual valuation space. By the monotonicity[3] of $\psi_1$, $a_{1p} = \psi_i^{-1}(\tilde{a}_{1p})$ is the corresponding threshold in $v$-space. Step 6-8 in the algorithm check for the condition that a slot $p$ is never allocated to bidder 1 but a more desirable slot $k < p$ is allocated at $a_{1k} < \infty$ and, in that case, set $a_{1p} = a_{1k}$ (This convention is implied in (13)). Step 10-12 use (13) to compute the prices $p_{1p}$ given the thresholds.

The dominating computation inside the FOR loops is the call to OPTMATCH, thus giving a $\mathcal{O}(n \cdot m^2 n) = \mathcal{O}(n^2 m^2)$ time complexity. ∎

The payment corresponding to the revenue maximizing allocation rule $\mathbf{X}^*$ can be computed efficiently using the Lemma 4 with $\psi_i(v_i) = \max(\nu_i(v_i), 0)$.

### 3.3 Rank based allocation rules

**Definition 1** (Rank based allocation rule). *A rank based allocation rule with rank vector $\mathbf{w} \in \mathbb{R}_+^n$ allocates slot $j$ to the bidder in the $j^{th}$ position in the decreasing order statistics of $\{w_i b_i\}_{i=1}^n$, where $\mathbf{b}$ is the vector of advertiser's bid.*

Let $\mathbf{X}^w$ denote the rank based allocation rule with ranking vector $\mathbf{w} \in \mathbb{R}_+^n$. Let $\boldsymbol{\gamma} = [w_1 b_1, \ldots, w_n b_n]$. Then under $\mathbf{X^w}$, the total click-through rate for bidder $i$ is given by

$$\sum_{j=1}^{m} c_{ij} X_{ij}^w(\mathbf{b}) = \sum_{j=1}^{m} (c_{ij} - c_{i,j+1}) \mathbf{1}(\gamma_i \geq \gamma_{[j]}^{-i}),$$

where $\gamma_{[j]}^{-i}$ denotes the j-th largest term in the vector $\boldsymbol{\gamma}_{-i}$ and $\mathbf{1}(\cdot)$ denotes the indicator function that takes the value 1 when its argument is true, and zero otherwise. Since $\gamma_i$ is increasing in $b_i$ and $c_{ij} \geq c_{i,j+1}$, it follows that the total click-through rate $\sum_{j=1}^{m} c_{ij} X_{ij}^w(\mathbf{b})$ is non-decreasing in $b_i$. Thus, by Lemma 2 part(a), $\mathbf{X}^w$ is incentive compatible.

---
[3]If $\psi_i(v_i)$ is flat in some interval $\psi_i^{-1}(.)$ is taken to right continuous at the point of discontinuity.

**Lemma 5.** *The unique per click price $p_{ij}^{\mathbf{w}}$, implementing the rank based allocation rule, $\mathbf{X}^{\mathbf{w}}$ are given by,*

$$p_{ij}^w(\mathbf{v}_{-i}) = \sum_{k=j}^{m} \left( \frac{c_{ik} - c_{i,k+1}}{c_{ij}} \right) \frac{\gamma_{[k+1]}}{w_i} \tag{18}$$

**Proof:** Fix $\mathbf{v}_{-i}$. The rank based allocation rule with ranking vector $\mathbf{w}$ allocates slot $j$ to advertiser $i$ iff $\gamma_{[j]} \geq w_i v_i > \gamma_{[j+1]}$. Thus, the thresholds $a_{ij}, j = 1, \ldots, m$ (see Lemma 2, part (b)) are equal to $\frac{\gamma_{[j+1]}}{w_i}$. Thus (18) follows from the alternative characterization of per click prices in Lemma 2, part (c). ∎

Aggarwal et al. (2006b) computes (18) using arguments especially tailored for rank based allocations.

The simplicity of the rank based mechanisms make them a very attractive. However, which rank vector $\mathbf{w}$ to use is far from clear! In particular, if the click-though-rate is not separable, i.e. $c_{ij} \neq \phi_i \mu_j$, then both the Google rank vector ($w_i = c_{i1}$) and the Yahoo! rank vector ($\mathbf{w} = \mathbf{1}$) neither maximize efficiency nor maximize revenue. We show that even when the click-through-rate is separable the revenue maximizing rank vector is *not* the same as the efficiency maximizing rank vector and the revenue maximizing mechanism is *not* rank-based!

**Example 2.** Consider an adword auction of two slots and two bidders with valuations, $v_i$ uniformly distributed on $[0, 1]$. Suppose a rank based auction mechanism awards the slot 1 to advertiser 1 if $v_1 \geq \alpha v_2$ and to merchant 2 otherwise. Define $A = c_{11} - c_{12}$ and $B = c_{21} - c_{22}$. Then the prices in (18) imply that the expected revenue of the auctioneer,

$$\Pi(\alpha) = \mathbb{E}_{(v_1, v_2)} \left[ A \alpha v_2 \mathbf{1}(v_1 \geq \alpha v_2) + B \frac{v_1}{\alpha} \mathbf{1}(v_2 \geq \frac{v_1}{\alpha}) \right]$$

Simplifying, we get

$$\Pi(\alpha) = \begin{cases} A \frac{1}{6\alpha} + B(\frac{1}{2\alpha} - \frac{1}{3\alpha^2}) & \text{if } \alpha \geq 1 \\ A(\frac{\alpha}{2} - \frac{\alpha^2}{3}) + B \frac{\alpha}{6} & \text{if } \alpha \leq 1 \end{cases}$$

Thus the optimum[4],

$$\alpha^* = \begin{cases} \frac{4B}{A+3B} & \text{if } B \geq A \\ \frac{3A+B}{4A} & \text{if } A \geq B \end{cases}$$

Similar calculation for efficiency shows that the expected social surplus $\mathbf{S}(\alpha)$ is given by,

$$\mathbf{S}(\alpha) = \begin{cases} \frac{1}{3\alpha} A - \frac{1}{6\alpha^2} B + \frac{1}{2}(c_{21} + c_{12}) & \text{if } \alpha \geq 1 \\ \frac{\alpha}{3} B - \frac{\alpha^2}{6} A + \frac{1}{2}(c_{22} + c_{11}) & \text{if } \alpha < 1 \end{cases}$$

and the efficiency maximizing rank vector $\alpha^e = \frac{B}{A}$. Thus, $\alpha^* \neq \alpha^e$.

---
[4]The optimal point can be graphically verified to be unique. For $\alpha \leq 1$, $\Pi(\alpha)$ is concave and for $\alpha \geq 1$, $\Pi(\alpha)$ even though neither convex nor concave, monotonically increases up to a constant $\min(1, \frac{3A+B}{4A})$ and decreases thereafter.

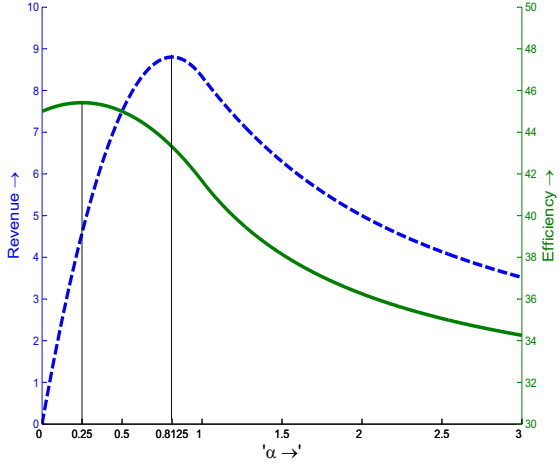

Figure 3: Revenue and efficiency as a function of $\alpha$ in Example 2

Now, assume that the click-through-rate is separable, $c_{ij} = \phi_i \mu_j$, for $\phi > 0$ and $\mu_1 \geq \mu_2 > 0$. Then the revenue maximizing rank vector

$$\alpha^* = \begin{cases} \frac{4\phi_2}{\phi_1 + 3\phi_2} & \text{if } \phi_2 \geq \phi_1 \\ \frac{3\phi_1 + \phi_2}{4\phi_1} & \text{if } \phi_1 \geq \phi_2 \end{cases}$$

and the efficiency maximizing rank vector $\alpha^e = \frac{\phi_2}{\phi_1}$. Since $v_i$ are uniformly distributed on $[0, 1]$, the virtual valuations $\nu_i(v_i) = v_i - \frac{1 - F_i(v_i)}{f_i(v_i)} = 2v_i - 1$, and the revenue maximizing allocation

$$\mathbf{X}^* \in \operatorname*{argmax}_{\mathbf{X} \in \mathcal{M}_{22}} \sum_{j=1}^{m} \mu_j \sum_{i=1}^{n} \phi_i (2v_i - 1) x_{ij}$$

Thus, $\mathbf{X}^*$ assigns slot 1 to bidder 1 if $v_1 \geq \frac{\phi_2}{\phi_1} v_2 + \frac{\phi_1 - \phi_2}{2\phi_1}$ and to bidder 2 otherwise. This is not a rank based allocation even though the click-through-rate is separable! The efficiency maximizing allocation rule is clearly rank based with $\alpha^e = \frac{\phi_2}{\phi_1}$.

Figure 3 plots the efficiency and revenue as a function of $\alpha$ for $\mathbf{c} = \begin{bmatrix} 50 & 10 \\ 50 & 40 \end{bmatrix}$. For this data, $\alpha^* = 0.8125$ and $\alpha^e = 0.25$. The efficiency maximizing ranking vector being more biased than the revenue maximizing vector because even though bidder 2 generates more value in slot 2, bidder 1 determine the slot price. ∎

We propose *customized rank based allocation rule* as an improvement over the rank based allocation rule. This rule reduces to the *top down* auction proposed in Aggarwal et al. (2006a) when $c_{ij} \equiv c_j$, i.e. bidder independent click-through-rates.

**Definition 2** (Customized Rank Based Allocation). *A customized rank based (CRB) allocation rule allocates slot $j$ to the bidder having highest order statistics of $\{c_{ij} b_i\}_{i=1}^{n}$ among those who have not been assigned a slot $< j$.*

**Algorithm 3** COMPUTECRPRICES

---
for $i = 1$ to $n$ do
   $\mathbf{S} \leftarrow \phi$, $\mathbf{N}_i \leftarrow \{1,\ldots,n\} \setminus \{i\}$, $j \leftarrow 1$, $c_{i,m+1} \leftarrow 0$, $a_{i,m+1} \leftarrow 0$.
   **for** $j = 1$ to $m$ **do**
     $(s, I) \leftarrow \max_{k \in \mathbf{N}_i \setminus \mathbf{S}}(\{c_{kj}v_i\})$.
     $a_{ij} \leftarrow \frac{s}{c_{ij}}$, $\mathbf{S} \leftarrow \mathbf{S} \bigcup I$
   **end for**
   **for** $j = 1$ to $m-1$ **do**
     **if** $a_{ij} < a_{i,j+1}$ **then**
       $a_{i,j+1} \leftarrow a_{ij}$
     **end if**
   **end for**
   **for** j=1 to m **do**
     $p_{ij} \leftarrow \frac{1}{c_{ij}} \sum_{k=j}^{m}(a_{ik} - a_{i,k-1})(c_{ij} - c_{i,k-1})$
   **end for**
end for

---

Using Lemma 2, it is straightforward to deduce that the CRB allocation is incentive compatible. Thus, the prices implementing CRB allocation rule satisfy (13). Algorithm COMPUTECRBPRICES computes the prices implementing customized rank based allocation rule in $\mathcal{O}(n^2m)$ time. Establishing correctness of the algorithm is straightforward and is left to the reader. Also, by using the virtual valuations,

$$\mathbf{z} = \left( \left(v_1 - \frac{1 - F_1(v_1)}{f_1(v_1)}\right)^+, \ldots, \left(v_n - \frac{1 - F_n(v_n)}{f_n(v_n)}\right)^+ \right)$$

instead of the valuations $\mathbf{v}$ in Algorithm COMPUTECRBPRICES, the customized rank based allocation rule can approximate revenue maximizing auction as well. In general, the **CRB** allocation rule is likely to out-perform any rank based allocation in terms of both efficiency and revenue generation.

### 3.4 Separable Click-though Rate

Suppose the click-through rate $c_{ij}$ is separable, i.e. $c_{ij} = \phi_i \mu_j$ with $\mu_1 \geq \mu_2 \geq \ldots \mu_m > 0$ and $\boldsymbol{\phi} > \mathbf{0}$. In this setting, the following results immediately follow from our results in the previous section.

1. The solution to the social (virtual) surplus maximization problem is rank based, i.e. the slot $j$ is awarded to the $j^{th}$ order statistics of $\{\phi_i v_i\}_{i=1}^n$ (respectively $\{\phi_i \nu_i(v_i)\}_{i=1}^n$).

2. The price-per-click given by (9) implementing the efficiency maximizing allocation are given

by

$$p_{[i]}^e(\mathbf{v}) = \frac{1}{\mu_i}(\mu_i - \mu_{i+1})\frac{\phi_{[i+1]}}{\phi_{[i]}}v_{[j+1]} + p_{[i+1]}^e$$

$$= \frac{1}{\mu_i}\sum_{j=i}^{m}(\mu_j - \mu_{j+1})\frac{\phi_{[j+1]}}{\phi_{[j]}}v_{[j+1]} \quad (19)$$

3. The price-per-click implementing revenue maximizing allocation rule is given by,

$$p_{[i]}^*(\mathbf{v}) = \frac{1}{\mu_i}\sum_{j=i}^{m}(\mu_j - \mu_{j+1})\nu_{[i]}^{-1}\left(\frac{\phi_{[j+1]}}{\phi_{[j]}}v_{[j+1]}\right) \quad (20)$$

where $\phi_{[k]}$ and $v_{[k]}$ represent the $\phi$ and bid of advertiser ranked $k$.

When the click-through-rate is separable, **CRB** allocation rule is the same as the Google allocation rule with rank vector $w_i = c_{i1}$ and is optimal for efficiency maximization.

## 4 Slotted Mechanism

In this section, we consider the per click valuations of the following form

$$v_{ij} = \begin{cases} v_i, & j \leq k_i \\ 0, & \text{otherwise,} \end{cases}$$

where the tuple $(v_i, k_i)$ is the private information (i.e. type) of the bidder $i$.

The efficiency maximization problem is given by

$$\begin{array}{ll} \max_{\mathbf{X} \in \mathcal{M}_{nm}} & \sum_{i=1}^{n}\sum_{j=1}^{m} v_i c_{ij} X_{ij} \\ \text{subject to} & X_{ij} = 0, \quad k_i < j \leq m, 1 \leq i \leq n, \end{array} \quad (21)$$

Since the a bid $\hat{k}_i > k_i$ is ex-post observable, the bidders are restricted to bid $\hat{k}_i \leq k_i$. Since advertisers have one-sided incentives about $k_i$, this is not a standard mechanism design problem. However, the specific structure of the problem, namely that for all $i$, $\mathbf{v} \in \mathbb{R}_+^n$ and $\mathbf{k}_{-i}$, the total surplus is non-decreasing in $k_i$, ensures incentive compatibility with respect to $k_i$ at the $VCG$ payments. When $k_i$ are common knowledge, the solution presented in § 3 can be directly applied by pointwise maximizing the virtual surplus.

The following Lemma characterizes the set of all incentive compatible allocation rules in the slotted valuation model. The proof is a simple adaptation of the proof of Lemma 1 in Iyengar and Kumar (2006).

**Lemma 6.** *The following are equivalent characterizations of IC allocation rules.*

(a) *The total click-through rate*

$$\sum_{j=1}^{k_i} c_{ij} X_{ij}\big((b, \mathbf{k}_i), (\mathbf{v}_{-i}, \mathbf{k}_{-i})\big)$$

*for each bidder $i$ is a non-decreasing function of $b$ for all $(\mathbf{v}_{-i}, \mathbf{k})$.*

(b) For all $i, \mathbf{v}_{-i}$ and $\mathbf{k}$, there exists thresholds $0 \leq a_{i,m} \leq a_{i,m-1} \leq \cdots \leq a_{i,k_i} \leq \infty$ such that bidder $i$ is allocated slot $j$ iff $v_i \in (a_{ij}, a_{i,j-1}]$.

(c) The utility of each bidder $i$ under allocation rule $\mathbf{X}$ is of the form

$$u_i(v_i, k_i; \mathbf{v}_{-i}, \mathbf{k}_{-i}) = \overline{u}_i(k_i, (\mathbf{v}_{-i}, \mathbf{k}_{-i})) + \int_0^{v_i} \Big( \sum_{j=1}^m c_{ij} X_{ij}((u, k_i), (\mathbf{v}_{-i}, \mathbf{k}_{-i})) \Big) du.$$

where $\overline{u}_i(k_i, (\mathbf{v}_{-i}, \mathbf{k}_{-i}))$ are non-negative and nondecreasing functions of $k_i$ appropriately chosen to ensure that $u_i(v_i, k_i; \mathbf{v}_{-i}, \mathbf{k}_{-i})$ is non-decreasing in $k_i$ for all $(\mathbf{v}, \mathbf{k}_{-i})$.

Lemma 6 part (a) and (c) implies that any feasible mechanism that is $IC$ with respect to the valuation bid can be made $IC$ with respect to the slot bid using independent side payments $\overline{u}_i(k_i, (\mathbf{v}_{-i}, \mathbf{k}_{-i}))$ so that the total surplus of each bidder is increasing with $k_i$. In the rest of this section, the term *nomimal surplus* will denote the term

$$\int_0^{v_i} \Big( \sum_{j=1}^m c_{ij} X_{ij}((u, k_i), (\mathbf{v}_{-i}, \mathbf{k}_{-i})) \Big) du.$$

Lemma 6, together a change in the order of integration (see Iyengar and Kumar (2006), Theorem 1), implies that the revenue maximization problem is equivalent to choosing a feasible allocation rule $\mathbf{X}$ with $X_{ij}(\mathbf{v}) = 0$ for all $k_i < j \leq m, 1 \leq i \leq n$ and a set of side payments $\overline{u}_i$ that maximize

$$\max_{\mathbf{X}, \overline{\mathbf{u}}} \mathbb{E}_{(\mathbf{v}, \mathbf{k})} \sum_{i=1}^n \left[ \sum_{j=1}^m c_{ij} \left( v_i - \frac{1 - F_i(v_i | k_i)}{f_i(v_i | k_i)} \right) X_{ij}(\mathbf{v}) - \overline{u}_i(k_i, \mathbf{v}_{-i}) \right]$$

subject to the constraint that $\sum_{j=1}^m c_{ij} X_{ij}(v_i, \mathbf{v}_{-i})$ is non-decreasing in $v_i$ for all $i, \mathbf{v}_{-i}, \mathbf{k}$, and $\overline{u}_i(k_i, \mathbf{v}_{-i}) + \int_0^{v_i} \Big( \sum_{j=1}^m c_{ij} X_{ij}((u, k_i), (\mathbf{v}_{-i}, \mathbf{k}_{-i})) \Big) du$ is non-decreasing with $k_i$ for all $i, \mathbf{v}, \mathbf{k}_{-i}$.

Since the social surplus depends on $k_i$, there exist bid profiles at which the total click through rate assigned to some bidder $i$ is not monotone in $k_i$. Hence, the nominal surplus,

$$u_i^0(\mathbf{v}, \mathbf{k}) \triangleq \int_0^{v_i} \Big( \sum_{j=1}^m c_{ij} X_{ij}((u, k_i), (\mathbf{v}_{-i}, \mathbf{k}_{-i})) \Big) du \qquad (22)$$

is not monotone in $k_i$. Consequently, there does not exist any regularity conditions on the prior distribution $f_i(v_i, k_i)$ that guarantees that the side payments $\overline{u}_i$ can be set equal to zero without loss of generality. Example 3 presents a concrete example to demonstrate this.

**Example 3.** Let $\mathbf{c} = \begin{bmatrix} 3 & 0 \\ 3 & 1 \end{bmatrix}$, and the realized $k_1 = k_2 = 2$. Suppose $v_i$ and $k_i$ are independent. The nominal surplus, given by (22), of bidder 2

$$u_2 = \begin{cases} 1(v_2 - 0) & \text{if } \hat{k}_2 = 2 \text{ and } v_2 < a_{21} = \nu_2^{-1}(\frac{3}{2}\nu_1(v_1)) \\ 3(v_2 - \hat{a}_{21}) & \text{if } \hat{k}_2 = 1 \text{ and } v_2 > \hat{a}_{21} = \nu_2^{-1}(\nu_1(v_1)) \end{cases}$$

Thus, if $(v_1, v_2)$ are such that $a_{21} > v_2 > \frac{3}{2}\hat{a}_{21}$, bidder 2 prefers to bids $\hat{k}_2 = 1$. Suppose the realized $\nu_1(v_1) = \frac{4}{17}$, then bidder 2 strictly prefers to bid $\hat{k}_2 = 1$ if $\frac{3}{2}\nu_2^{-1}\left(\frac{4}{17}\right) < v_2 < \nu_2^{-1}\left(\frac{6}{17}\right)$.

Suppose the CDF of $v_2$ is given by

$$F_2(x) = 3x\mathbf{1}_{\{(0,\frac{1}{4})\}}(x) + \left[\frac{3}{4} + \frac{1}{4}\left(1 - e^{-\frac{1}{12}(x-\frac{1}{4})}\right)\right]\mathbf{1}_{\{[\frac{1}{4},\infty)\}}(x)$$

i.e. a $\left(\frac{3}{4}, \frac{1}{4}\right)$ mixture of a uniform distribution on $\left(0, \frac{1}{4}\right)$ and an exponential distribution with rate $\frac{1}{12}$ on $[\frac{1}{4}, \infty)$ respectively. Then the corresponding virtual valuation function is

$$\nu_2(x) = \left(2x - \frac{1}{3}\right)\mathbf{1}_{\{0,\frac{1}{4}\}}(x) + \left(x - \frac{1}{12}\right)\mathbf{1}_{\{\frac{1}{4},\infty\}}(x)$$

so that $\nu_2^{-1}(\frac{4}{17}) = \frac{29}{102}$ and $\nu_2^{-1}(\frac{6}{17}) = \frac{89}{204}$. Thus for all $v_2 \in \left(\frac{29}{68}, \frac{89}{204}\right)$ bidder 2 strictly prefers to bid $\hat{k}_2 = 1$.

Consequently pointwise optimal solution with side payments $\overline{u}$ set to zero does not induce truth-telling and, hence, is sub-optimal in this example. ∎

Vohra and Malakhov (2005) only considers the $IC$ mechanisms with $\overline{u} = 0$. The above example shows that this constraint might eliminate a number of reasonable allocation rules.

Lemma 6 part(b) and a simple adaptation of Lemma 4 gives the following result.

**Lemma 7.** *Given any set of $\psi_i : \mathbb{R}_+ \mapsto \mathbb{R}_+$, $\psi_i \in \mathcal{C}[\mathbb{R}]$ and $\psi_i$ non-decreasing, the nominal surplus,*

$$\int_0^{v_i} \left(\sum_{j=1}^m c_{ij} X_{ij}^\psi((u, k_i), (\mathbf{v}_{-i}, \mathbf{k}_{-i}))\right) du = \sum_{j=1}^m \left[c_{ij}v_i - \sum_{k=j}^m (a_{ik} - a_{i,k+1})(c_{i,j} - c_{i,k+1})\right] X_{ij}^\psi(\mathbf{v}, \mathbf{k})$$

*can be computed in $\mathcal{O}(m^2 n^2)$ for any solution to (21) with $v_i$ replaced by $\psi_i(v_i)$.*

The proof of this lemma is left to the reader.

Next, we discuss two heuristic mechanisms for maximizing revenue. The first heuristic mechanism is defined as follows. If there exists intervals over which the virtual valuations $\nu_i(v_i, k_i) = v_i - \frac{1-F_i(v_i|k_i)}{f_i(v_i|k_i)}$ is increasing in $v_i$, use the ironing procedure detailed in (section 3.2) Iyengar and Kumar (2006) to compute ironed out virtual valuation $\hat{\nu}_i(v_i, k_i)$ that is non-decreasing in $v_i$. Set the allocation rule

$$\mathbf{X}(\mathbf{v}, \mathbf{k}) \in \operatorname{argmax} \sum_{i=1}^n \sum_{j=1}^m c_{ij}\hat{\nu}_i(v_i, k_i) X_{ij}(\mathbf{v}). \tag{23}$$

Since $\hat{\nu}_i$ is non-decreasing in $v_i$, $\mathbf{X}(\mathbf{v}, \mathbf{k})$ is $IC$. Set the transfer payment

$$T_i(\mathbf{v}, \mathbf{k}) = \sum_{j=1}^{k_i} c_{ij}v_i X_{ij}(\mathbf{v}, \mathbf{k}) - u_i^0(\mathbf{v}, \mathbf{k}) - \overline{u}_i(\mathbf{v}, k_i) \tag{24}$$

where

$$\overline{u}_i(\mathbf{v}, \mathbf{k}) = \max_{\hat{k}_i \leq k_i}\left[u_i^0((v_i, \hat{k}_i), \mathbf{t}_{-i}) - u_i^0((v_i, k_i), \mathbf{t}_{-i})\right] \tag{25}$$

This payment ensures that the total surplus of each bidder $i$ is non-decreasing in $k_i$. Since side payments are identically zero in the $VCG$ mechanism, it follows that the side payment above would be identically zero if the virtual valuation are replaced by valuation. Lemma 7 implies that this mechanism can be implemented in $\mathcal{O}(m^3 n^2)$ time.

Second mechanism is **CRB** as described in §3 applied to the slotted valuation model, i.e., for $j = 1, \ldots, m$ allocate the slot $j$ to a bidder having highest order statistics of $\{c_{ij} \nu_i(v_i, k_i)\}_{i=1}^n$ among those who have not been assigned a slot number less than $j$ and who have $k_i \leq j$.

Assume that for all $i$, $v_i$, the virtual valuation function $\nu_i$ is non-decreasing in $v_i$ and $k_i$. Since, for all $i$, $\mathbf{v}_{-i}$ and $\mathbf{k}$, the click-through-rate of the slot allocated to advertiser $i$ is non-decreasing in $v_i$, this mechanism is IC. Also note that for all $i$, $\mathbf{v}$, $\mathbf{k}_{-i}$, the click-through-rate of the slot allocated to advertiser $i$ is non-decreasing in $k_i$, which implies that the nominal surplus, $u_i^0(\mathbf{v}, \mathbf{k})$ is non-decreasing in $k_i$. Thus, we can set the side payments $\overline{u}_i$ to be identically zero for this allocation rule, i.e. under **CRB**, the advertisers does not earn any information rent due to the $k_i$ dimension of the bid. Moreover, algorithm COMPUTECRBPRICES can be used $k_i$ times for each $i$ to compute the prices that implement the slotted **CRB** mechanism in $\mathcal{O}(m^3 n)$ time. The following lemma summarize the above results.

**Lemma 8.** *If the virtual valuations $\nu_i(v_i, k_i), i = 1, \ldots, n$ are non-decreasing in $v_i$ and $k_i$ then $X_{CRB}^*$ is incentive compatible with unique minimal prices and side payments $\overline{u}_i$ identically set to zero. Furthermore, these prices can be computed in $\mathcal{O}(m^3 n)$ time.*

Note that, given an $IC$ allocation rule, $\mathbf{X}$ the nominal expected revenue,

$$\mathbb{E}_{(\mathbf{v},\mathbf{k})}\Big[\sum_{i=1}^n \sum_{j=1}^{k_i} c_{ij} v_i X_{ij}(\mathbf{v}) - u_i^0(\mathbf{v}, \mathbf{k})\Big] \tag{26}$$

is an upper bound on the optimal expected revenue achievable by $\mathbf{X}$. Since pointwise maximum $\mathbf{X}^*$ defined in (23) is an upper bound on the optimal expected revenues, setting $\mathbf{X} = \mathbf{X}^*$ in (26) gives an upper bound on the achievable optimal expected revenue. We compare the relative performance of the two sub-optimal mechanisms proposed in the numerical study presented in the next section, which is remarkable for the synthetic data set used.

Finally, we remark that the problem of finding optimal mechanism is hard because we need to select both an allocation rule and a set of side payments independently. This is because the transfer payment implementing a given IC allocation rule is not uniquely defined by the rule. We conclude this section with the following result.

**Lemma 9.** *Suppose the click-through-rates are separable. Then the CRB mechanism is optimal for social surplus maximization. When the virtual surplus is non-decreasing in both $v_i$ and $k_i$, the CRB with $\nu(v_i)$ is optimal for revenue maximization.*

## 5 Comparing the mechanisms: a numerical study

In this section, we report the results of a numerical study that compares the revenue and efficiency properties of the **RB** and **CRB** mechanisms as compared to the optimal mechanisms in the slot independent valuation model and the slotted model. We consider an adword auction model with 4 slots and 6 bidders, i.e. $m = 4, n = 6,$. The click-through-rate matrix is given by

$$\mathbf{C} = \begin{bmatrix} 96 & 93 & 47 & 42 \\ 90 & 75 & 24 & 3 \\ 83 & 62 & 19 & 7 \\ 50 & 45 & 42 & 36 \\ 95 & 90 & 82 & 63 \\ 93 & 80 & 77 & 2 \end{bmatrix}$$

where $c_{ij}$ denotes the expected number of clicks per day when bidder $i$ is assigned to slot $j$. We take the common random number approach and compare the average performance on a sample of $N = 10,000$ valuation vectors. For each bidder $i = 1, \ldots, 6$, the synthetic valuations are generated from a gamma distribution with mean 5 and standard deviation $\sqrt{5}$, i.e. $v_i \sim \text{GAMMA}(5,1)$.

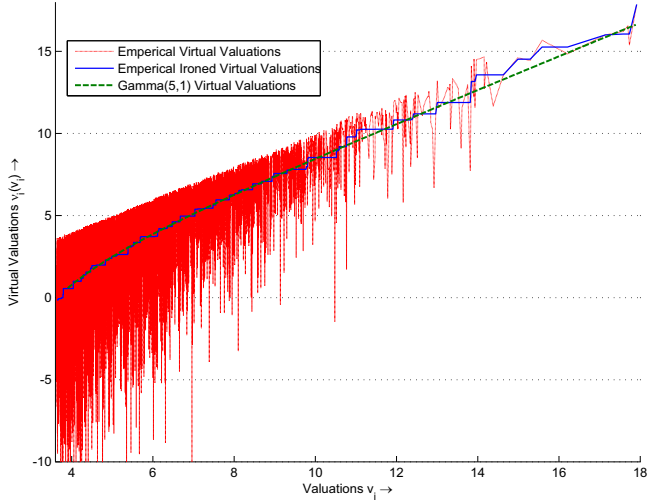

Figure 4: Ironed empirical virtual valuations vs. GAMMA(5,1) virtual valuations

For virtual surplus maximization, we used the virtual valuations $\nu_i^{\text{GAMMA}}$ defined by the GAMMA(5,1) distribution[5] as well as the virtual valuation $\tilde{\nu}_i$ implied by the samples. Let $v_i^{[t]}$ denote the value in the $t$-th position when the samples $\{v_i^{(t)}\}_{t=1}^N$ are sorted in increasing order. Then

$$\tilde{\nu}_i(v_i^{[t]}) = v_i^{[t]} - (v_i^{[t+1]} - v_i^{[t]})\frac{1 - \frac{t}{N}}{\frac{1}{N}} \quad t = 1, \ldots, N-1, \tilde{\nu}_i(v_i^{[N]}) = v_i^{[N]}$$

Since the empirical virtual valuations is not monotone, we iron it using the ironing procedure in Myerson (1981) to get monotone ironed empirical virtual valuations $\hat{\nu}_i$. From Figure 4 that displays $\tilde{\nu}_i, \hat{\nu}_i$ and $\nu_i^{\text{GAMMA}}$ in the range of valuation where they are positive it is clear that all these three quantities are very close.

---

[5] $\nu_i^{\text{GAMMA}}$ is increasing since Gamma(5,1) is strictly log-concave.

## 5.1 Results for the uniform slot valuation model

As observed in Example 2 computing the optimal expected revenue (efficiency) maximizing rank vectors is not a convex optimization problem and, therefore, cannot be solved efficiently. We circumvent this issue by taking the common random number approach and maximizing the average revenue (or efficiency) over the sampled data. For example, revenue maximizing rank vector

$$\mathbf{w}^* \in \operatorname*{argmax}_{\mathbf{w} \in \mathbb{R}^n_+, w_1 = 1} \frac{1}{N} \sum_{t=1}^{N} \left\{ \sum_{i=1}^{m} \sum_{j=i}^{m} (c_{[j]^{(t)}, j} - c_{[j]^{(t)}, j+1}) \frac{\gamma^{(t)}_{[j+1]^{(t)}}}{w^{(t)}_{[j]}} \right\},$$

where $\gamma^{(t)} = (w_1^* v_1^{(t)}, \ldots, w_n^* v_n^{(t)})$, $\mathbf{v}^{(t)}$ denotes the valuation vector for the $t$-th sample, and $[j]^{(t)}$ denotes the index of bidder ranked at the $j$-th position in the $t$-sample. For a given a ranking vector and sample $\mathbf{v}^{(t)}$, the slot prices can be computed in $\mathcal{O}(n \log(n) + m^2)$ time; therefore, one can attempt optimizing the rank vector $\mathbf{w}$ using a derivative-free NLP method. We use a Matlab based non-linear optimizer to solve these problems. Since the optimization problem is unconstrained and low dimensional, the Matlab code is quite stable and converges quickly. The efficiency maximizing rank vector was also computed from the samples.

Table 1 displays the average revenue and efficiency for the following mechanisms.

1. The revenue maximizing **RB** mechanism that uses the sample-based optimal rank vector $\mathbf{w}^*$.

2. The efficiency maximizing **RB** that uses the sample-based optimal rank vector $\mathbf{w}^e$.

3. heuristic **RB** mechanism: this mechanism uses the Google rank vector $w_i = \frac{c_{i,1}}{c_{1,1}}$ for efficiency maximization and Yahoo rank vector (i.e. $\mathbf{w} = \mathbf{1}$) for revenue maximization.

4. The **CRB** mechanism: for efficiency maximization the **CRB** ranks using the valuations $v_i$ and for revenue maximization the **CRB** ranks using the virtual valuations $\nu^{\textsc{Gamma}}(v_i)$ and $\hat{\tilde{\nu}}_i(v_i)$.

5. The optimal mechanism for the true prior $\textsc{Gamma}(5, 1)$ and the empirical prior distribution.

The superscripts "$*$" and "$e$" denote the revenue maximizing mechanisms and efficiency maximizing mechanism respectively. In the cells corresponding to the **CRB** and the optimal mechanism in Table 1, the number inside (resp. outside) the bracket denotes the value obtained by using the virtual valuation $\hat{\nu}_i$ (resp. $\nu_i^{\textsc{Gamma}}$). For the optimal **RB** mechanisms the revenue and efficiency was computed using the same set of $N$ samples that were used to compute the rank vectors.

The optimal **RB** achieves 91.95% (92.24%) of the optimal revenue – in contrast, the **CRB** rule achieves 95.75% (95.71%), thus providing a 3.80% (3.47%) improvement in revenues. Similarly, optimal **RB** achieves 87.53% of the optimal efficiency and the **CRB** achieves 93.15% of the optimal efficiency; thus, improving efficiency by 5.62% over the optimal **RB**.

Table 2 displays the optimal rank vectors, average slot prices and the advertiser's surplus for optimal **RB** and heuristic **RB** under both efficiency maximizing and revenue maximizing mechanisms. These numerical results support the following observations.

a) The efficiency maximizing rank vector is more *biased* (i.e. away from **1**) as compared to the revenue maximizing rank vector. This is consistent with what was observed earlier in Example 2.

|  | Revenue Maximization | | Efficiency Maximization | |
|---|---|---|---|---|
|  | $\Pi^*$ | $S^*$ | $\Pi^e$ | $S^e$ |
| Heurestic **RB** | 998.24 | 1523.91 | 938.70 | 1463.47 |
| Optimal **RB** | 1020.30 | 1558.90 | 1002.90 | 1571.40 |
| **CRB** | 1062.45(1058.62) | 1585.85(1595.81) | 829.36 | 1672.30 |
| Optimal | 1109.58(1106.08) | 1687.53(1697.96) | 1000.93 | 1795.24 |

Table 1: The revenue and efficiency in **RB**, **CRB** and optimal mechanisms

| | Optimal **RB** | | | | | | Heuristic **RB** | | | |
|---|---|---|---|---|---|---|---|---|---|---|
| | R.M. | | | E.M. | | | R.M. | | E.M. | |
| | $w^*$ | $u^*$ | $p^*$ | $w^e$ | $u^e$ | $p^e$ | $u^*$ | $p^*$ | $u^e$ | $p^e$ |
| 1 | 1 | 105.19 | 4.5549 | 1 | 110.17 | 4.3125 | 140.09 | 4.3294 | 106.55 | 4.5954 |
| 2 | 1.1413 | 92.29 | 4.3256 | 1.2949 | 123.12 | 4.1794 | 91.73 | 4.1312 | 67.22 | 4.3611 |
| 3 | 1.1420 | 80.74 | 3.9090 | 1.0524 | 66.43 | 3.7919 | 68.28 | 3.7273 | 59.22 | 3.9414 |
| 4 | 0.8196 | 44.65 | 3.5485 | 0.6457 | 24.23 | 3.4902 | 71.54 | 3.4333 | 69.64 | 3.5342 |
| 5 | 0.9395 | 115.47 | | 0.9538 | 126.82 | | 32.81 | | 131.50 | |
| 6 | 1.0462 | 100.24 | | 1.1093 | 117.67 | | 120.34 | | 91.54 | |

Table 2: The average slot prices and advertiser surplus in optimal **RB** and heuristic **RB** mechanisms

b) The **CRB** mechanism tailored for revenue maximization results in higher average prices for each slot and is, therefore, able to extract higher surplus.

c) The **CRB** mechanism tailored for efficiency maximization results in in low slot prices and, consequently, low revenues. Thus, it is important that we maximize the virtual surplus when **CRB** is used for revenue maximization.

d) The revenue generated by the **RB** mechanisms when maximizing surplus and virtual surplus is comparable indicating that **RB** is a robust mechanism.

## 5.2 Results for slotted model

We assumed that the privately known slot value $k_i \sim \text{unif}\{1, \ldots, 4\}$ and generated $N = 10000$ samples. We used the same set of samples for the valuation as in the previous subsection.

Table 4 displays the average revenue and efficiency with the pointwise maximizer allocation rule and **CRB** allocation rule. Table 5 displays the average advertiser surplus, average slot prices and average side payments to each bidder. Note that the total average side payment is remarkably low 6.40 (3.3237) when compared to the average revenue. The data in the table together with the bound (26) implies that 988.61+6.40=995.01 (984.9536+3.3237= 988.2763) is an upper bound on the achievable revenue. The pointwise maximizer achieves 99.36% (99.66%) revenue of this bound, while the **CRB** achieves 94.78% (95.05%) revenue of this upper bound. Thus, both proposed mechanisms are likely to be close to optimal. For efficiency, **CRB** achieves 92.79% of the true optimal $VCG$ (or pointwise maximizer with $\nu_i(v_i) = v_i$) efficiency as shown in Table 4.

|      | **CRB**         |                 |        |        | Optimal         |                 |        |        |
|------|-----------------|-----------------|--------|--------|-----------------|-----------------|--------|--------|
|      | R.M.            |                 | E.M.   |        | R.M.            |                 | E.M.   |        |
|      | $u^*$           | $p^*$           | $u^e$  | $p^e$  | $u^*$           | $p^*$           | $u^e$  | $p^e$  |
| 1    | 121.93(125.53)  | 4.7052(4.6609)  | 185.92 | 3.8542 | 115.89(119.26)  | 4.4409(4.4139)  | 181.92 | 4.1223 |
| 2    | 61.26(62.58)    | 4.3955(4.3441)  | 56.19  | 3.3017 | 97.09(99.69)    | 4.1567(4.1124)  | 114.42 | 3.2994 |
| 3    | 48.23(46.41)    | 3.8203(3.7871)  | 30.09  | 2.2039 | 86.80(84.74)    | 3.7102(3.6775)  | 81.86  | 2.8008 |
| 4    | 58.45(62.89)    | 3.2472(3.2937)  | 115.06 | 1.6034 | 51.31(55.22)    | 2.9123(2.9355)  | 52.82  | 2.9273 |
| 5    | 130.23(133.05)  |                 | 280.58 |        | 117.65(119.83)  |                 | 194.21 |        |
| 6    | 103.31(106.72)  |                 | 175.12 |        | 109.22(113.14)  |                 | 169.08 |        |

Table 3: The average slot prices and advertiser surplus in **CRB** and optimal mechanisms

|                     | Revenue Maximization |                   | Efficiency Maximization |         |
|---------------------|---------------------|-------------------|------------------------|---------|
|                     | $\Pi^*$             | $S^*$             | $\Pi^e$                | $S^e$   |
| Pointwise Maximizer | 988.61 (984.95)     | 1468.91(1475.55)  | 806.50                 | 1562.49 |
| **CRB**             | 943.11 (939.37)     | 1385.20 (1392.49) | 695.64                 | 1449.87 |

Table 4: The revenue and efficiency in **CRB** and optimal mechanisms

Table 5 presents the average side payments, average advertiser surplus and the average slot prices per click for both the mechanisms. The prices tabulated are not adjusted for side payments. This is because side payments could be positive even when no slot is allocated to the advertiser (this happens when at a particular bid profile the advertiser can get a positive allocation by underbidding $k_i$; however, no slot is allocated at the true $k_i$). We found that the distribution of side payments is very skewed, i.e. even though the average side payments are very small, at certain bid profiles the side payments are of the same order of magnitude as the advertiser surplus.

|   | Pointwise Maximizer |                 |                   |        |        | **CRB**        |                |        |        |
|---|---------------------|-----------------|-------------------|--------|--------|----------------|----------------|--------|--------|
|   | R.M.                |                 |                   | E.M.   |        | R.M.           |                | E.M.   |        |
|   | $u^*$               | $p^*$           | $\overline{u}^*$  | $u^e$  | $p^e$  | $u^*$          | $p^*$          | $u^e$  | $p^e$  |
| 1 | 97.68(100.15)       | 4.9157(4.8965)  | 1.7401(0.9592)    | 164.73 | 4.6090 | 114.61(118.04) | 5.0799(5.0405) | 196.07 | 4.2767 |
| 2 | 78.55(80.70)        | 4.3183(4.2910)  | 0.8961(0.6066)    | 101.03 | 3.4710 | 61.77(63.17)   | 4.243(4.2082)  | 73.19  | 3.0186 |
| 3 | 66.60(65.06)        | 3.1226(3.1033)  | 0.6893(0.5558)    | 69.42  | 1.6009 | 48.47(46.51)   | 2.7492(2.7484) | 46.36  | 1.1265 |
| 4 | 37.36(39.61)        | 1.0297(1.0457)  | 0.4743(0.1776)    | 58.29  | 0.2571 | 35.84(38.51)   | 1.0054(1.0260) | 71.52  | 0.0888 |
| 5 | 101.78(103.31)      |                 | 1.6600(0.5830)    | 195.57 |        | 97.02(99.04)   |                | 207.09 |        |
| 6 | 91.93(95.12)        |                 | 0.9417(0.4405)    | 166.95 |        | 84.38(87.86)   |                | 160.00 |        |

Table 5: The average slot prices and advertiser surplus in **CRB** and optimal mechanisms in the slotted mechanism

# 6   Conclusion

The discrete structure of the adword auction allocation space allows us to reduce the $IC$ constraint to the existence of bidder dependent slot prices at which each bidder self-selects their allocated slot. We use this new characterization to show that in this auction models there are $IC$ mechanisms which are not affine maximizers. Achieving optimal revenues with multi-dimensional types in this model remains a hard stochastic program involving multi-dimensional ironing and sweeping.

In slot independent private valuation model, the pricing characterization collapses to the existence of ordered threshold at which a given advertiser is allocated particular slots. The prices implementing any $IC$ allocation rule can by computed very efficiently. When the click through rate is not separable, we show that the revenue (efficiency) maximizing rank vector can be efficiently computed using the history of bids available for each adword. When prior information is unavailable, the proposed **CRB** allocation rule is a very attractive choice since it is simple, non-parametric, computationally efficient and has a superior performance to any rank based mechanism. Our numerical study also indicate that both **RB** and **CRB** achieve close to optimal revenues and efficiency without using detailed prior information.

In the slotted auction model even though the set of $IC$ mechanism is easy to characterize, computing the optimal mechanism problem is a hard because of the non-zero side payments needed to screen the slot information. Our numerical study indicates that the two sub-optimal mechanisms proposed in this paper are likely to perform close to optimal.

# References

Aggarwal, G., Feldman, J., and Muthukrishnan, S. (2006a). Bidding to the top: Vcg and equilibria of position-based auctions. Working Paper.

Aggarwal, G., Goel, A., and Motvani, R. (2006b). Truthful auctions for pricing search keywords. In *Proceedings of the 2006 ACM Conference on Electronic Commerce*, Ann Arbor, MI.

Ahuja, R. K., Magnanti, T., and Orlin, J. (1993). *Network Flows: Theory, Algorithms and Applications*. Prentice Hall, NJ.

Bikhchandani, S. and Ostroy, J. M. (2006). From the assignment model to combinatorial auctions. In Peter Cramton, Y. S. and Steinberg, R., editors, *Combinatorial Auctions*. MIT Press, Cambridge, MA.

Chung, K.-S. and Ely, J. C. (2002). Ex-post incentive compatible mechanisms. Working Paper, Department of Economics, Northwestern University.

Edelman, B., Ostrovsky, M., and Schwarz, M. (2005). Internet advertising and the generalized second price auction: Selling billions of dollars worth of keywords. NBER Working paper 11765.

Feng, J. (2006a). Optimal mechanism for selling a set of commonly ranked objects. Working Paper.

Feng, J. (2006b). Price cycles in online advertising auctions. Working Paper.

Feng, J., Bhargava, H. K., and Pennock, D. M. (2005). Implementing sponsored search in web search engines: Computational evaluation of alternative mechanisms. Fothcoming in *Informs Journal on Computing*.

Groves, T. (1979). Efficient collective choice with compensation. In Laffont, J.-J., editor, *Aggregation and Revelation of Preferences*, pages 37–59. Amsterdam: North-Holland.

Hongwei, G., Muller, R., and Vohra, R. V. (2004). Dominant strategy mechanisms with multidimensional types. Working Paper.

Iyengar, G. and Kumar, A. (2006). Optimal procurement auctions for divisible goods with capacitated suppliers. Technical report. CORC Tech Report - TR-2006-01, Available at http://www.corc.ieor.columbia.edu/reports/techreports/tr-2006-01.pdf.

Kitts, B., Laxminarayan, P., LeBlanc, B., and Meech, R. (2005). A formal analysis of search auctions including predictions on click fraud and bidding tactics. In *Workshop on Sponsored Search Auctions, ACM Electronic Commerce*, Vancouver, British Columbia, Canada.

Lahaie, S. (2006). An analysis of alternative slot auction designs for sponsored search. In *Proceedings of the 2006 ACM Conference on Electronic Commerce*, Ann Arbor, MI.

Lavi, R., Mu'alem, A., and Nisan, N. (2004). Towards a characterization of truthful combinatorial auctions. Working Paper.

Leonard, H. (1983). Elicitation of honest preferences for the assignment of individuals to positions. *Journal of Political Economy*, 91(3).

Lim, W. S. and Tang, C. S. (2004). An auction model arising from an internet search service provider. Fothcoming in *European Journal of Operations Research*.

Liu, D. and Chen, J. (2005). Designing online auctions with past performance information. Working Paper.

Milgrom, P. R. (2004). *Putting Auction Theory to Work*. Cambridge University Press, Cambridge, UK.

Myerson, R. B. (1981). Optimal auction design. *Mathematics of Operations Research*, 6(1):58–73.

Roberts, K. (1979). The characterization of implementable choice rules. In Laffont, J.-J., editor, *Aggregation and Revelation of Preferences*, pages 321–348. Amsterdam: North-Holland.

Rochet, J. C. (1987). A condition for rationalizability in a quasi-linear context. *Journal of Mathematical Economics*, 16:191–200.

Rochet, J.-C. and Chone, P. (1998). Ironing, sweeping, and multidimensional screening. *Econometrica*, (4):783–826.

Saks, M. and Yu, L. (2005). Weak monotonicity suffices for truthfulness on convex domains. Working Paper, Available at http://www.math.rutgers.edu/~saks/PUBS/truthful.ecsub.pdf.

Shapley, L. and Shubik, M. (1972). The assignment game-I: the core. *Inter-national Journal of Game Theory*, 1:111–130.

Varian, H. R. (2006). Position auctions. working paper.

Vohra, R. V. and Malakhov, A. (2005). Optimal auction for capacitated bidders - a network approach. Working Paper, Managerial Economics and Decision Sciences, Kellogg School of Management, Northwestern University.

Zhan, R. L., Shen, Z.-J. M., and Fengz, J. (2005). Ranked items auctions and online advertisement. Working Paper.